\documentclass[a4paper,aps,pra,twocolumn,floatfix,superscriptaddress,notitlepage,usenames,dvipsnames,svgnames,table,nofootinbib,longbibliography]{revtex4-2}
\pdfoutput=1
\usepackage[utf8]{inputenc}
\usepackage[T1]{fontenc}
\usepackage{graphicx}
\usepackage{grffile}
\usepackage{wrapfig}
\usepackage{rotating}
\usepackage[normalem]{ulem}
\usepackage{amsmath}
\usepackage{textcomp}
\usepackage{amssymb}
\usepackage{capt-of}
\usepackage{soul}
\usepackage{parskip}
\usepackage{mathrsfs}
\usepackage[margin= 0.75in]{geometry}
\usepackage[braket, qm]{qcircuit}
\usepackage{footmisc}
\usepackage{pgfplots}

\definecolor{blue-violet}{rgb}{0.54, 0.17, 0.89}
\usepackage{hyperref}
\hypersetup{
 pdfauthor={Faidon Andreadakis, Emanuel Dallas, and Paolo Zanardi},
 pdftitle={Long-time Quantum Scrambling and Generalized Tensor Product Structures},
 pdflang={English},colorlinks=true,linkcolor=RubineRed,citecolor=blue-violet,urlcolor=Cerulean} 
\usepackage[capitalise]{cleveref}

\usepackage[english]{babel}
\usepackage{letltxmacro}
\LetLtxMacro{\ORIGselectlanguage}{\selectlanguage}
\makeatletter
\DeclareRobustCommand{\selectlanguage}[1]{%
  \@ifundefined{alias@\string#1}
    {\ORIGselectlanguage{#1}}
    {\begingroup\edef\x{\endgroup
      \noexpand\ORIGselectlanguage{\@nameuse{alias@#1}}}\x}%
}
\newcommand{\definelanguagealias}[2]{%
  \@namedef{alias@#1}{#2}%
}
\makeatother
\definelanguagealias{en}{english}
\definelanguagealias{eng}{english}

\usepackage{bbm}
\usepackage{etoolbox}
\usepackage{tcolorbox} 
\usepackage{tikz}
\usetikzlibrary{patterns}
\usepackage{amsthm}
\usepackage{amssymb}
\usetikzlibrary{arrows.meta}
\usepackage{mathtools}

\usepackage{bm}
\usepackage{dsfont}
\usepackage[font=small,labelfont=bf,justification=RaggedRight,format=plain]{caption}
\usepackage{subcaption}
\usepackage{enumerate}
\usepackage{hhline}

\usepackage{thmtools}
\usepackage{thm-restate}

\newtheorem*{conj}{Conjecture}
\newtheorem*{principle}{Principle}
\newtheorem{definition}{Definition}
\newtheorem{proposition}{Proposition}

\DeclareMathOperator{\diag}{diag}

\usepackage{microtype}
\usepackage{dsfont}
\usepackage{tensor}

\usepackage{physics}
\usepackage{mathrsfs}
\usepackage{mathtools}
\usepackage{fixltx2e}
\usepackage{relsize}

\DeclarePairedDelimiterX\phys[2]{\langle}{\rangle}{#1 \delimsize\vert\mathopen{} #2}

\theoremstyle{remark}
\newtheorem{exmp}{Example}

\begin{document}

\title{Long-time Quantum Scrambling and {Generalized Tensor Product Structures}}

\author{Faidon Andreadakis}
\email [e-mail: ]{fandread@usc.edu}
\affiliation{Department of Physics and Astronomy, and Center for Quantum Information Science and Technology, University of Southern California, Los Angeles, California 90089-0484, USA}

\author{Emanuel Dallas}
\email [e-mail: ]{dallas@usc.edu}
\affiliation{Department of Physics and Astronomy, and Center for Quantum Information Science and Technology, University of Southern California, Los Angeles, California 90089-0484, USA}

\author{Paolo Zanardi}
\email [e-mail: ]{zanardi@usc.edu}
\affiliation{Department of Physics and Astronomy, and Center for Quantum Information Science and Technology, University of Southern California, Los Angeles, California 90089-0484, USA}
\affiliation{Department of Mathematics, University of Southern California, Los Angeles, California 90089-2532, USA}

\date{\today}

\begin{abstract}

{Much recent work has been devoted to the study of information scrambling in quantum systems. In this paper, we study the long-time properties of the algebraic out-of-time-order-correlator (``$\mathcal{A}$-OTOC'') and derive an analytical expression for its long-time average under the non-resonance condition. The $\mathcal{A}$-OTOC quantifies quantum scrambling with respect to degrees of freedom described by an operator subalgebra $\mathcal{A}$, which is associated with a partitioning of the corresponding system into a generalized tensor product structure. Recently, the short-time growth of the $\mathcal{A}$-OTOC was proposed as a criterion to determine which partition arises naturally from the system's unitary dynamics. In this paper, we extend this program to the long-time regime where the long-time average of the $\mathcal{A}$-OTOC serves as the metric of subsystem emergence.} Under this framework, natural system partitions are characterized by the tendency to minimally scramble information over long time scales. We consider several physical examples{, ranging from quantum many-body systems and stabilizer codes to quantum reference frames,} and perform {the} minimization of {the $\mathcal{A}$-OTOC long-time average} both analytically and numerically over relevant families of algebras. For simple cases subject to the non-resonant condition, minimal $\mathcal{A}$-OTOC long-time average is shown to be related to minimal entanglement of the Hamiltonian eigenstates across the emergent system partition. Finally, we conjecture and provide evidence for a general structure of the algebra that minimizes the average for non-resonant Hamiltonians. 
\end{abstract}
\maketitle

\section{Introduction} \label{secintro}
\par Information-theoretic properties of quantum dynamics provide insights applicable to a wide range of physical systems. Quantum information scrambling is a prominent such property and refers to the dynamical generation of correlations among initially distinguishable degrees of freedom. A wide range of phenomena, from thermalization in quantum many-body systems to black hole physics and holography \cite{larkin_quasiclassical_1969,noauthor_alexei_nodate,maldacena_bound_2016,lashkari_towards_2013,polchinski_spectrum_2016,mezei_entanglement_2017,roberts_chaos_2017,hayden_black_2007,shenker_black_2014}, have been linked to scrambling dynamics, which are commonly studied using the out-of-time-order correlator (OTOC) as a diagnostic tool \cite{roberts_diagnosing_2015,swingle_unscrambling_2018,xu_scrambling_2022,mi_information_2021,braumuller_probing_2022,li_measuring_2017,nie_experimental_2020,garttner_measuring_2017,chen_detecting_2020,landsman_verified_2019}. 

\par {In this paper, we investigate scrambling dynamics in the long-time regime. This regime is relevant when the timescale of interest far exceeds the characteristic quantum (i.e., “microscopic”) system timescale and has been examined in studies of quantum chaos \cite{rammensee_many-body_2018,kidd_saddle-point_2021,fortes_gauging_2019} and quantum phase transitions \cite{wang_probing_2019}. We use an algebraic-out-of-time-order correlator ($\mathcal{A}$-OTOC) as a metric of information scrambling between an algebra of observables and its commutant under dynamics, which provides a unified framework that incorporates operator entanglement \cite{zanardi_entanglement_2001,yan_information_2020,styliaris_information_2021} and coherence-generating power \cite{zanardi_coherence-generating_2017,zanardi_measures_2017} as special cases. As correlation functions do not have infinite-time limits under unitary dynamics in finite-dimensional systems, we probe the long-time behavior of the $\mathcal{A}$-OTOC via its long-time average (LTA). Notably, for the case of bipartite algebras, the scaling of the LTA distinguishes between the chaotic and integrable phases of quantum many-body systems \cite{styliaris_information_2021,anand_brotocs_2021}.}
\par {The $\mathcal{A}$-OTOC LTA becomes more analytically tractable under the assumption that the system dynamics satisfy the no-resonance condition (NRC), meaning the energy spectrum contains no degeneracies nor degenerate gaps. The condition of nondegenerate energy gaps is generically satisfied by fully interacting Hamiltonians \cite{linden_quantum_2009}, while the full NRC is expected to hold, either exactly or approximately, for generic chaotic Hamiltonians \cite{haake_quantum_2010}. Furthermore, the addition of small random perturbations generally lifts any degeneracies so that the NRC is satisfied, even though the effect of such perturbations may be significant only in large time scales.}

\par Given a specific background structure (e.g., in terms of spatial locality, such as a collection of neighboring qubits) and physical system of interest, suitable averages of OTOCs over the corresponding degrees of freedom quantify its scrambling dynamics and reveal connections with other information-theoretic concepts, such as operator entanglement and entropy production \cite{styliaris_information_2021,zanardi_information_2021,anand_brotocs_2021}, quantum coherence \cite{anand_quantum_2021}, quasiprobabilities \cite{yunger_halpern_quasiprobability_2018}, and more \cite{yan_information_2020,touil_information_2021,borgonovi_timescales_2019}.
\par However, a reversed approach can be taken: given a dynamics, different partitionings of the system can be distinguished by their distinct scrambling properties, revealing an emergent structure. This corresponds to an information-theoretic approach to quantum \textit{mereology} \footnote{The term mereology is borrowed from philosophy, where it has a rich history. For a succinct overview, see \cite{varzi_mereology_2019}.}, the study of parthood relations in quantum systems \cite{carroll_quantum_2021,zanardi_operational_2023,nager_taxonomy_2021}.
\par Recently, Zanardi et al. \cite{zanardi_operational_2023} used the short-time expansion of {the $\mathcal{A}$-OTOC} as a mereological diagnostic criterion. The algebraic approach to quantum mereology relies on two general mathematical properties of Hilbert spaces. First, Hilbert spaces are not equipped with unique tensor product structures, even if a physical system may have a “natural” decomposition (e.g., a collection of qubits, or an obvious system and bath). Second, a subalgebra of observables induces a specific decomposition of its Hilbert space\footnote{See \cref{sec:preliminaries} for a formal statement.}. This “structural” freedom was recognized in previous work on virtual quantum subsystems \cite{zanardi_virtual_2001,zanardi_quantum_2004} and is relevant to decoherence-free subspaces \cite{zanardi_noiseless_1997,lidar_decoherence-free_1998}, noiseless subsystems \cite{lidar_quantum_2013,zanardi_stabilizing_2000}, operator-error correction \cite{kribs_unified_2005} and quantum reference frames \cite{bartlett_reference_2007}. It is intuitive, as Zanardi et al. proposed, to consider the minimization of the short-time $\mathcal{A}$-OTOC as a selection of a preferred Hilbert space decomposition into parts which scramble their informational identities most slowly.
\par {Using our results for the $\mathcal{A}$-OTOC LTA, we extend this framework to the long-time limit. We interpret the minimization of the LTA over algebras as a \textit{selection} of a preferred decomposition of a system at ``macroscopic'' timescales. This selection is a task of fundamental interest in understanding emergent structure and phenomena in quantum systems, as well as comparing the differences in structure between short and long time scales. It may also serve as a practical diagnostic for studying scrambling of operationally accessible observables on quantum circuits \cite{mi_information_2021}. In some cases, the $\mathcal{A}$-OTOC minimization can be interpreted as \textit{fixing} an algebra of observables and identifying a least-scrambling unitary dynamics, which may be a more meaningful interpretation in examples involving quantum control.} 
\par In \cref{sec:preliminaries}, we formally introduce the $\mathcal{A}$-OTOC as well as the decomposition of the Hilbert space induced by an algebra and its commutant. In \cref{alg_lta}, we take the long-time average of the $\mathcal{A}$-OTOC and derive expressions for the $\mathcal{A}$-OTOC LTA that hold in non-resonant systems, showcasing their use numerically for a family of stabilizer algebras. In \cref{sec:minimization} we study the minimization of the LTA for unitary families of NRC Hamiltonians, providing a simple analytical application in the context of quantum reference frames, as well as exhibiting numerically the connection of the bipartite $\mathcal{A}$-OTOC LTA with the average eigenstate mutual information in certain quantum many-body systems. Lastly, \cref{conclusion} provides conclusions and steps forward for future research.

\section{Preliminaries}
\label{sec:preliminaries}
\par Consider a finite-dimensional quantum system represented by a Hilbert space $\mathcal{H}\cong\mathbb{C}^d$. Any physical observable is represented by a linear operator and we denote as $\mathcal{L}(\mathcal{H})$ the space of all linear operators on $\mathcal{H}$. $\mathcal{L}(\mathcal{H})$ is also a Hilbert space equipped with the Hilbert-Schmidt inner product: $\langle X,Y\rangle=\Tr(X^\dagger \, Y)$. For closed quantum systems in the Heisenberg picture, the time evolution of a physical observable $X\in\mathcal{L}(\mathcal{H})$ is given as $\mathcal{U}_t (X) = U_t \, X U_t^\dagger$, where $U_t = \exp{it \, H}$ is the unitary evolution generated by the system Hamiltonian $H$.
\par The central mathematical structures in this paper are hermitian-closed, unital subalgebras $\mathcal{A}\subset\mathcal{L}(\mathcal{H})$ that are used to describe the relevant degrees of freedom of interest. The symmetries of $\mathcal{A}$ constitute the commutant algebra $\mathcal{A}^\prime=\{Y\in \mathcal{A}^\prime \, \lvert \, [Y,X]=0 \; \forall X \in \mathcal{A}\}$ and correspond to degrees of freedom initially uncorrelated with $\mathcal{A}$. Due to the double commutant theorem, $(\mathcal{A}^\prime)^\prime = \mathcal{A}$ \cite{davidson_c-algebras_1996}, and therefore these algebras can be considered as pairs $(\mathcal{A},\mathcal{A}^\prime)$. Under time evolution, information is scrambled between $\mathcal{A}$ and $\mathcal{A}^\prime$, quantified by the $\mathcal{A}$-OTOC \cite{andreadakis_scrambling_2023}:
\begin{definition} \label{AOTOC}
The $\mathcal{A}$-OTOC of algebra $\mathcal{A}$ and unitary $\mathcal{U}_t$ is defined as: 
\begin{equation} \label{def_aotoc}
    G_{\mathcal{A}}(\mathcal{U}_t)=\frac{1}{2d}{\mathlarger{\mathbb{E}}}_{X_\mathcal{A},Y_{\mathcal{A}^\prime}} \left[ \left\lVert X_{\mathcal{A}}, \mathcal{U}_t(Y_{\mathcal{A}^\prime})\right\rVert_2^2 \right].
\end{equation}

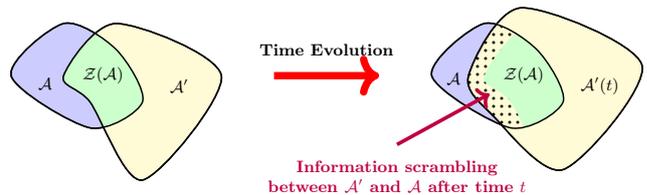
\begin{figure}
    \centering
     \resizebox{0.5\textwidth}{!}{
    \begin{tikzpicture}
  \draw[thick, fill=blue!20] plot[smooth cycle] coordinates{(0,0) (1,1) (2,0.5) (2.5,-0.5) (1.5,-1)};
  
  \draw[thick, fill=yellow!20] plot[smooth cycle] coordinates{(1,0) (2,1) (4,0) (2.5,-2) (1.5,-0.5)};
  
  \begin{scope}
    \clip plot[smooth cycle] coordinates{(0,0) (1,1) (2,0.5) (2.5,-0.5) (1.5,-1)};
    \fill[green!20] plot[smooth cycle] coordinates{(1,0) (2,1) (4,0) (2.5,-2) (1.5,-0.5)};
  \end{scope}

    \draw[thick] plot[smooth cycle] coordinates{(0,0) (1,1) (2,0.5) (2.5,-0.5) (1.5,-1)};
    \draw[thick] plot[smooth cycle] coordinates{(1,0) (2,1) (4,0) (2.5,-2) (1.5,-0.5)};

  \node at (0.65, -0.15) {$\mathcal{A}$};
  \node at (1.75, 0) {$\mathcal{Z(A)}$};
  \node at (3.2, -0.2) {$\mathcal{A'}$};

  \draw[->, line width=4pt, red] (5,0) -- (7,0) node[midway, above=6pt, black, font=\bfseries] {Time Evolution};
  

\begin{scope}[shift = {(8,0)}]
  \draw[thick, fill=blue!20] plot[smooth cycle] coordinates{(0,0) (1,1) (2,0.5) (2.5,-0.5) (1.5,-1)};
  \draw[thick, fill=yellow!20] plot[smooth cycle] coordinates{(0.7,0) (1.7,1.3) (4,0.2) (2.5,-1.8) (1.15,-0.75)};
  \begin{scope}
    \clip plot[smooth cycle] coordinates{(0,0) (1,1) (2,0.5) (2.5,-0.5) (1.5,-1)};
    \draw[pattern=dots, pattern color = black!95] plot[smooth cycle] coordinates{(0.7,0) (1.7,1.3) (4,0.2) (2.5,-1.8) (1.15,-0.75)};
    \draw[purple,dashed] plot[smooth cycle] coordinates{(0.7,0) (1.7,1.3) (4,0.2) (2.5,-1.8) (1.15,-0.75)};
  \end{scope}
    
  \begin{scope}
    \clip plot[smooth cycle] coordinates{(0,0) (1,1) (2,0.5) (2.5,-0.5) (1.5,-1)};
    \fill[green!20] plot[smooth cycle] coordinates{(1,0) (2,1) (4,0) (2.5,-2) (1.5,-0.5)};
  \end{scope}

    \draw[thick] plot[smooth cycle] coordinates{(0,0) (1,1) (2,0.5) (2.5,-0.5) (1.5,-1)};
    \draw[thick] plot[smooth cycle] coordinates{(0.7,0) (1.7,1.3) (4,0.2) (2.5,-1.8) (1.15,-0.75)};

  \node at (0.42, -0.05) {$\mathcal{A}$};
  \node at (1.75, 0) {$\mathcal{Z(A)}$};
  \node at (3.2, -0.2) {$\mathcal{A'}(t)$};

   \draw[->, line width=2pt, purple] (-0.66,-1.32) -- (1.1,-0.35) node[pos=0, below=5pt, purple, font=\bfseries, align=center] {Information scrambling \\ between $\mathcal{A'}$ and $\mathcal{A}$ after time $t$};
   
\end{scope}

\end{tikzpicture}
}
    \caption{A visual representation of information scrambling induced by Hamiltonian time evolution. The ``component'' of $\mathcal{A'}(t)$ that ``leaks'' into $\mathcal{A}$ corresponds to the scrambling; this is represented by the dotted component of $\mathcal{A'}(t)$.}
    \label{fig:scrambling_picture}
\end{figure}

\end{definition}
$\mathbb{E}_{X_\mathcal{A},Y_{\mathcal{A}^\prime}}$ denotes the Haar average over the unitaries $X_\mathcal{A} \in \mathcal{A}$ and $Y_{\mathcal{A}^\prime}\in\mathcal{A}^\prime$. The evolution of operators in $\mathcal{A}^\prime$ under $\mathcal{U}_t$ leads to potential non-commutativity with operators in $\mathcal{A}$, which is intuitively interpreted as information scrambling between the corresponding degrees of freedom. This is operator spreading in the space of algebras, where locality is defined relative to $\mathcal{A}$. The $\mathcal{A}$-OTOC thus provides an information scrambling measure with respect to a generalized locality structure (see \cref{structure_H}), independent of a specific choice of operators $X_\mathcal{A}$ and $Y_{\mathcal{A}^\prime}$.
\par Importantly, the subalgebra of observables $\mathcal{A}$ induces a decomposition of the Hilbert space $\mathcal{H}$ into a direct sum of virtual quantum subsystems \cite{zanardi_virtual_2001,zanardi_quantum_2004}, referred to as generalized tensor product structure (gTPS) \cite{zanardi_operational_2023}. Specifically, denoting the center of the algebra as $\mathcal{Z}(\mathcal{A})\coloneqq \mathcal{A} \cap \mathcal{A}^\prime$ with dimension $\dim{\mathcal{Z}(\mathcal{A})}=d_\mathcal{Z}$, we have
\begin{equation} \label{structure_H}
    \mathcal{H} = \bigoplus_{J=1}^{d_Z} \mathcal{H}_J, \quad \mathcal{H}_J \cong \mathbb{C}^{n_J} \otimes \mathbb{C}^{d_J}.
\end{equation}
where $\mathcal{A}$ and $\mathcal{A}^\prime$ act irreducibly on the $\mathbb{C}^{d_J}$ and $\mathbb{C}^{n_J}$ factors respectively:
\begin{equation} \label{structure_A}
    \mathcal{A} \cong \bigoplus_{J=1}^{d_Z} \mathbb{I}_{n_J}\otimes \mathcal{L}(\mathbb{C}^{d_J}), \quad \mathcal{A}' \cong \bigoplus_{J=1}^{d_Z} \mathcal{L}(\mathbb{C}^{n_J})\otimes\mathbb{I}_{d_J}. 
\end{equation}
From \cref{structure_H,structure_A}, we have $d=\sum_J n_J d_J$, $\dim{\mathcal{A}} \coloneqq d(\mathcal{A}) = \sum_J d_J^2$ and $\dim{\mathcal{A}^\prime} \coloneqq d(\mathcal{A}') = \sum_J n_J^2$.
\par For any given algebra $\mathcal{A}$, there is an orthogonal completely positive projection map $\mathbb{P}_{\mathcal{A}}: \mathcal{L}(\mathcal{H}) \rightarrow \mathcal{A}$, such that $\mathbb{P}_\mathcal{A}=\mathbb{P}_\mathcal{A}^2$, $\mathbb{P}_\mathcal{A}=\mathbb{P}_\mathcal{A}^\dagger$, $\Im \mathbb{P}_\mathcal{A}=\mathcal{A}$. In terms of \cref{structure_A}, we have $\mathbb{P}_\mathcal{A} (\cdot )=\oplus_J \frac{\mathds{1}_{n_J}}{n_J} \otimes \Tr_{n_J}(\cdot )$ and, similarly, $\mathbb{P}_{\mathcal{A}^\prime} (\cdot )=\oplus_J \Tr_{d_J}(\cdot )\otimes \frac{\mathds{1}_{d_J}}{d_J}$. Using appropriate orthogonal bases $\{f_\gamma\}_{\gamma=1}^{d(\mathcal{A}^\prime)}$ of $\mathcal{A}^\prime$ and $\{e_\alpha\}_{\alpha=1}^{d(\mathcal{A})}$ of $\mathcal{A}$, the projection maps have the Kraus representations $\mathbb{P}_\mathcal{A} (\cdot )=\sum_\gamma f_\gamma (\cdot ) f_\gamma^\dagger$, $\mathbb{P}_{\mathcal{A}^\prime} (\cdot ) = \sum_\alpha e_\alpha (\cdot ) e_\alpha^\dagger$ (see \cref{bases_formula} for an explicit expression).

\section{Algebra scrambling in the long-time limit}\label{alg_lta}
We begin with a Hamiltonian with a spectral decomposition $H=\sum_{k=1}^M E_k \Pi_k$, where $M$ is the number of distinct energy levels. The scrambling properties of the unitary dynamics $U_t = \exp{it \, H}$ over algebras $\mathcal{A}$ in the long-time regime can be quantified by utilizing the infinite-time average of the $\mathcal{A}$-OTOC, hereafter referred to as $\mathcal{A}$-OTOC long-time time average (LTA):
\begin{equation} \label{lta_exact}
\overline{G_{\mathcal{A}}(\mathcal{U}_t)}^t \coloneqq \lim_{T\rightarrow \infty} \frac{1}{T} \int_{0}^T G_{\mathcal{A}}(\mathcal{U}_t) \, dt.
\end{equation}
Typical quantum systems reach equilibrium \cite{gogolin_equilibration_2016,short_quantum_2012,reimann_equilibration_2012}, whence the $\mathcal{A}$-OTOC relaxes to an equilibration value given by the LTA.

\subsection{THE $\mathcal{A}$-OTOC UNDER THE NRC}

We now look to the non-resonant regime. There are two types of systems that we consider. The first are those satisfying the NRC\textsuperscript{+} \cite{styliaris_information_2021}: energy gaps are non-degenerate but there exist energy level degeneracies. The second are those with both non-degenerate energy gaps and energy levels, satisfying the full NRC. The LTA for each is written below:
\begin{proposition} \label{prop_NRC}
    \begin{enumerate}[i.]
        \item \begin{equation} \label{NRCp}
        \begin{split}
        &\overline{G_{\mathcal{A}}(\mathcal{U}_t)}^{NRC^+} =\\
        &1 - \frac{1}{d}\left[ \sum\limits_{\gamma} \left(\lVert \mathbb{P}_{\mathcal{A}^\prime}(D_H (f_{\gamma}))\rVert_2^2 - \frac{1}{2}\sum\limits_k \lVert \mathbb{P}_{\mathcal{A}^\prime}(\Pi_k f_{\gamma} \Pi_k)\rVert_2^2 \right) +\right. \\ 
        &\quad\quad+\left.\sum\limits_{\alpha}\left(\lVert \mathbb{P}_{\mathcal{A}}(D_H (e_{\alpha}))\rVert_2^2 - \frac{1}{2}\sum\limits_k \lVert \mathbb{P}_{\mathcal{A}}(\Pi_k e_{\alpha} \Pi_k)\rVert_2^2 \right)\right],
        \end{split}
        \end{equation}
        where $D_H (\cdot )\coloneqq \sum_{k=1}^M \Pi_k (\cdot ) \Pi_k$ is the dephasing map with respect to the eigenprojectors of $H$ and $f_\gamma$, $e_\alpha$ are given in \cref{bases_formula}.

        \item \begin{equation} \label{NRC}
        \begin{split}
        \overline{G_{\mathcal{A}}(\mathcal{U}_t)}^{NRC} = 1-\frac{1}{d}&\left(\sum_{\mathcal{X}=\{\mathcal{A},\mathcal{A}^\prime\}}\Tr( R^{(0),\mathcal{X}}R^{(1),\mathcal{X}^\prime} ) - \right.\\
        &\left.\quad\quad -\frac{1}{2}\Tr( R_D^{(0),\mathcal{X}}R_D^{(1),\mathcal{X}^\prime} )\right),
        \end{split}
        \end{equation}
        where, for algebra $\mathcal{X}$, we define $R_{lk}^{(0),\mathcal{X}} \coloneqq \left\lVert \mathbb{P}_{\mathcal{X}}\left(\ket{\phi_k}\bra{\phi_l} \right) \right\rVert_2^2$, $R_{kl}^{(1),\mathcal{X}} \coloneqq  \left\langle\mathbb{P}_{\mathcal{X}}\left(\Pi_k\right),  \mathbb{P}_{\mathcal{X}}\left(\Pi_l\right)\right\rangle$, and for matrix $M$, $M_D \coloneqq diag(M)$.

    \end{enumerate}

\end{proposition}

A few observations follow directly. First, the infinite-time average ensures that these NRC LTAs are independent of any specific energy eigenvalues. Any Hamiltonians sharing a fixed set of eigenstates and satisfying the NRC will have the same $\mathcal{A}$-OTOC LTA. Second, the above NRC expressions are generally \textit{not} bounds on the exact LTA for systems not satisfying the NRC --- the net sum of terms discarded in these expressions can be either positive or negative. A notable exception to this is for the important special case of algebras that satisfy $n_J = \lambda \, d_J \; \forall \, J$ for some integer $\lambda$, whence the $\mathcal{A}$-OTOC has a neat geometrical representation and the NRC approximation provides an upper bound \cite{zanardi_quantum_2022}. Finally, systems with energy degeneracies do not necessarily have a unique NRC LTA. This is because the NRC expression depends on a choice of eigenbasis within degenerate energy levels, since all projectors in the formula are one-dimensional. These last two observations are especially relevant for usage of the NRC LTA expression as an approximation for the exact value.

\begin{figure*}
\centering
\begin{subfigure}{.5\textwidth}
  \centering
  \includegraphics[width=1\linewidth]{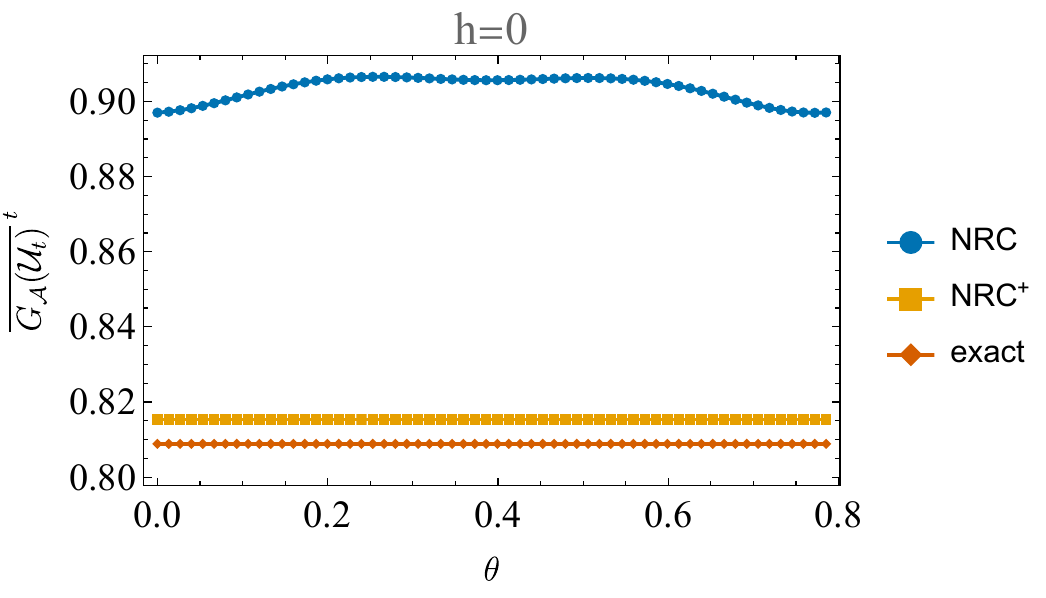}
 \caption{}\label{fig_minima}
\end{subfigure}%
\begin{subfigure}{.5\textwidth}
  \centering
  \includegraphics[width=1\linewidth]{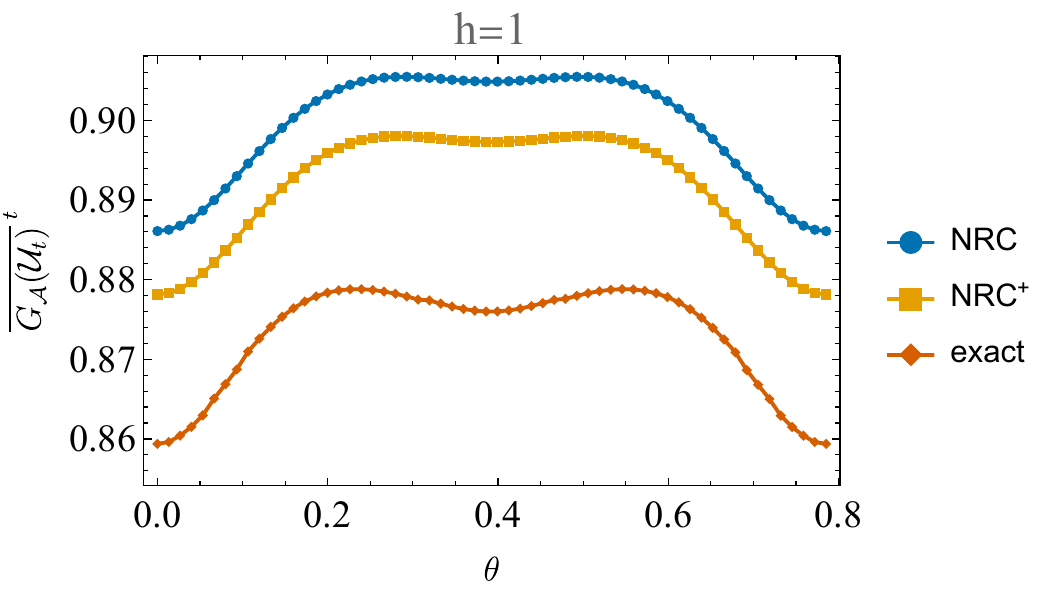}
  \caption{}\label{fig_funmin}
\end{subfigure}
\caption{The $\mathcal{A}$-OTOC LTA for the family of stabilizer algebras $\mathscr{A}=\{\exp(i\theta \sigma_y) \mathcal{A_{\text{5}}} \exp(-i\theta \sigma_y) \, \vert \, \theta \in [0,\pi/4] \}$ of the 5-qubit ``perfect'' code with the dynamics given by the Heisenberg model. (a) For $h=0$ the exact $LTA$ is constant for all the stabilizer algebras since the rotation $\exp{i\theta\sigma_y}$ is a symmetry of the Hamiltonian; the NRC\textsuperscript{+} approximation accurately captures this (since the eigenprojectors are similarly invariant), however, the NRC approximation gives a non-constant value as the chosen eigenbasis is not necessarily invariant under the rotation. (b) For $h=1$ both the NRC and NRC\textsuperscript{+} approximation accurately capture the behavior of the exact LTA, with minima occuring at $\theta =0$ and $\theta = \pi/4$.}

\label{fig_comp}
\end{figure*}

\begin{exmp}[\textbf{Stabilizer algebras}]
Let $G_S$ be the stabilizer group generated by the stabilizer operators $\{S_l \}_{l=1}^{n-k}$ associated with a stabilizer code with $k$ logical qubits. Let $\mathcal{A}_{st} \coloneqq \mathbf{C}[G_S]$ be the group algebra of $G_S$. Then, the Hilbert space decomposes into $2^k$-dimensional sectors $\mathcal{H} \cong \oplus_{J=1}^{2^{n-k}} \mathbb{C}^{2^k}$ that correspond to the encoding and syndrome subspaces. In terms of \cref{structure_H}, we have $d_J = 1$ and $n_J = 2^k \, \forall \, J=1,\dots, 2^{n-k}$.
\par Suppose that we have a parametrized family of such stabilizer algebras $\{\mathcal{A}_{st}(\theta )\}_\theta$ and some unitary dynamics $\mathcal{U}$. The behavior of the $\mathcal{A}$-OTOC LTA as a function of $\theta$ describes the variation of the (on average) scrambling of the corresponding encoding and syndrome subspaces by the dynamics in long time scales. In general, determining this behavior for a given family of algebras is non-trivial, but we expect that in many cases the \cref{NRCp,NRC} may serve as ``proxies'' to the exact behavior of $\overline{G_{\mathcal{A}}(\mathcal{U}_t)}^t$. To illustrate this consider the family of algebras $\mathscr{A}=\{\exp(i\theta \sigma_y) \mathcal{A_{\text{5}}} \exp(-i\theta \sigma_y) \, \vert \, \theta \in [0,\pi/4] \}$, where $\mathcal{A_{\text{5}}}$ is the stabilizer algebra associated to the stabilizer operators of the 5-qubit ``perfect'' code: $\{XZZXI,IXZZX,XIXZZ,ZXIXZ\}$. For the unitary dynamics, we consider the Heisenberg model with closed boundary conditions and a magnetic field term $H=\sum_{i=1}^5 \left(h\, \sigma_z^i + \vec{\sigma}^i \vec{\sigma}^{i+1} \right)$, where $\vec{\sigma}^6 \equiv \vec{\sigma}^1$ and $h$ is a coupling constant.
\par Using this family of algebras and dynamics, we compute as a function of $\theta$ the NRC and NRC\textsuperscript{+} approximations of the $\mathcal{A}$-OTOC LTA, as well as the exact one obtained by numerical simulations of the time evolution. We perform the above computations for $h=0$ and $h=1$ (see \cref{fig_comp}). For $h=1$, the NRC and NRC\textsuperscript{+} approximations accurately capture the behavior of the exact LTA, with minima occurring at $\theta=0$ and $\theta=\pi/4$, which corresponds to a rotation of $X\rightarrow Z$, $Z\rightarrow -X$. On the other hand, for $h=0$, the rotation $\exp(i\theta\sigma_y)$ is a symmetry of the Hamiltonian, so the exact LTA is constant as a function of $\theta$; this behavior is accurately captured by the NRC\textsuperscript{+}, whereas the NRC approximation gives a non-constant value since the chosen (non-unique) eigenbasis is not necessarily invariant under the Hamiltonian symmetry.
\end{exmp}

\section{$\mathcal{A}$-OTOC long-time-average Minimization}\label{sec:minimization}
In this section, we will be interested in using \cref{NRCp,NRC} to minimize the $\mathcal{A}$-OTOC LTA over families of algebras. A main motivation for this endeavor comes from Ref. \cite{zanardi_operational_2023}, where the short-time expansion of the $\mathcal{A}$-OTOC, referred to as the ``Gaussian scrambling rate'' $\tau_S^{-1}$, was introduced by Zanardi et al. as a mereological criterion for any algebra $\mathcal{A}$. Specifically, it was shown that $G_{\mathcal{A}}(\mathcal{U}_t)=2(t/\tau_S)^2+O(t^3)$, where
\begin{equation} \label{gaus}
    \tau_S^{-1}=\left\lVert \left(1-\mathbb{P}_{\mathcal{A}+\mathcal{A}^\prime}\right) \frac{H}{\sqrt{d}} \right\rVert_2
\end{equation}
At short time scales, this quantity describes the rate at which information scrambles from $\mathcal{A}$ to its commutant $\mathcal{A'}$ and is related to the distance of the Hamiltonian $H$ from the ``no-scrambling'' operator subspace $\mathcal{A}+\mathcal{A}^{\prime}$. Given the association between algebras and virtual partitions of systems (\cref{structure_H}), the Gaussian scrambling rate describes how quickly a collection of parts loses its ``informational identity.'' Equipped with this interpretation, Zanardi et al. argued that it is natural to consider partitions which retain their identity the longest to be dynamically ``preferred.'' Thus, for a family of algebras $\left\{\mathcal{A_\theta}\right\}_\theta$ and Hamiltonian $H$, the algebra $\mathcal{A}_{min} \in \left\{\mathcal{A_\theta}\right\}_\theta$ with the smallest $\tau_S^{-1}$ ``emerges'' as the dynamically preferred decomposition of the system under the action of $H$.
\par
In this paper, we follow the same general framework as Ref. \cite{zanardi_operational_2023} but focus on long time scales. We use the $\mathcal{A}$-OTOC LTA as our mereological criterion, which we interpret as measuring the extent to which a decomposition of the Hilbert space persists, on average, with informational ``integrity.'' The remainder of this paper thus relies on the following:
\begin{principle}\hypertarget{principle_1}{}
    Given a Hamiltonian $H$, and set of Hilbert space partitions represented by a parametrized family of algebras $\left\{\mathcal{A_\theta}\right\}_\theta$, the dynamically preferred partition at long time scales, $\mathcal{A}_{min} \in \left\{\mathcal{A_\theta}\right\}_\theta$, is the one that minimizes $\overline{G_{\mathcal{A}}(\mathcal{U}_t)}^t$.
\end{principle}

\par We can utilize \cref{NRC} to analytically minimize the $\mathcal{A}$-OTOC LTA over unitary families of NRC Hamiltonians. Note that, in general, for \cref{def_aotoc}, $G_\mathcal{A}(\mathcal{W}\,\mathcal{U}_t \, \mathcal{W}^\dagger)=G_{\mathcal{W}^\dagger(\mathcal{A})}(\mathcal{U}_t )$, where $\mathcal{W}$ is some unitary channel. This means that optimizing over unitary families of Hamiltonians is dual to optimizing over unitary families of algebras.
\begin{proposition} \label{nrc_thm}
Let $\mathcal{A}\cong \oplus_{J=1}^{d_\mathcal{Z}} \mathds{1}_{n_J} \otimes \mathcal{L}(\mathbb{C}^{d_J} )$ be an algebra of observables and $H_f$ be a family of NRC Hamiltonians that respect the superselection structure of the gTPS induced by $\mathcal{A}$, i.e. $H\in H_f \Leftrightarrow H=\oplus_J H_J$ and $H$ satisfies NRC. Then:
\begin{enumerate}[i.]
\item The Hamiltonian with eigenstates that are product states over the virtual quantum subsystems, i.e. $\{\ket{\psi_J}\otimes\ket{\phi_J} \, \vert \, J=1,\dots d_\mathcal{Z}; \, \psi_J = 1,\dots, n_J; \, \phi_J = 1,\dots , d_J \}$, minimizes the $\mathcal{A}$-OTOC LTA NRC. The minimum value is 
\begin{equation} \label{NRC_min}
    \overline{G_{\mathcal{A}}}^{NRC}_{min}=1-\frac{1}{d} \left( \sum_J d_J + \sum_J n_J - d_\mathcal{Z} \right)
\end{equation}
\item In particular, if $\mathcal{A}$ is a bipartite algebra ($d_\mathcal{Z}=1$), the Hamiltonian with product eigenstates over the bipartition minimizes the bipartite OTOC LTA NRC over all NRC Hamiltonians. The minimum value is 
\begin{equation}
    \overline{G}^{NRC}_{min}=1-\frac{1}{d_1}-\frac{1}{d_2}+\frac{1}{d_1d_2}
\end{equation}
where $\mathcal{H}\cong \mathbb{C}^{d_1}\otimes\mathbb{C}^{d_2}$.
\end{enumerate}
\end{proposition}
The above proposition shows that for NRC Hamiltonians that are block diagonal with respect to the gTPS (\cref{structure_H}) induced by $\mathcal{A}$, the minimum of the $\mathcal{A}$-OTOC LTA NRC is achieved exactly by Hamiltonians with eigenstates that have zero entanglement across the virtual quantum subsystems of the gTPS. Note that these eigenstates are unique up to local unitaries on each virtual bipartition. 
\par In addition, if $\mathcal{A}$ or $\mathcal{A}^\prime$ is abelian (i.e. if $n_J=1 \, \forall \, J$ or $d_J=1 \, \forall \, J$),  there is always an NRC block diagonal Hamiltonian that belongs in $\mathcal{A}^\prime$ or $\mathcal{A}$, respectively. Indeed, the NRC minimum of $\cref{NRC_min}$ is $\overline{G_{\mathcal{A}}}^{NRC}_{min}=0$, when there is no information scrambling. A special case is when $\mathcal{A}$ is a maximally Abelian algebra $\{\ket{k}\bra{k}\}_{k=1}^d$, whence the $\mathcal{A}$-OTOC is equal to the coherence-generating power of $U$ in the basis $B=\{\ket{k}\}_{k=1}^d$\cite{zanardi_quantum_2022,andreadakis_scrambling_2023}. In this case, minimal scrambling corresponds to Hamiltonians that have $B$ as the eigenbasis.
\par Based on the above results, we conjecture that, for a given algebra $\mathcal{A}$, the NRC Hamiltonians with eigenstates of the form $\{\ket{\psi_J}\otimes\ket{\phi_J} \, \vert \, J=1,\dots d_\mathcal{Z}; \, \psi_J = 1,\dots, n_J; \, \phi_J = 1,\dots , d_J \}$ minimize the $\mathcal{A}$-OTOC LTA over the set of all NRC Hamiltonians. Note that the conjectured Hamiltonian is diagonal in the basis\footnote{The distinguished basis corresponding to $\mathcal{A}$ consists of vectors $\ket{p_J}\otimes\ket{k_J}$ $(p_J=1,\dots n_J, \, k_J=1,\dots,d_J)$ and is unique up to local unitaries in each $J$-block. \label{distinguished_basis}} naturally selected by the algebra $\mathcal{A}$. Intuitively, such an eigenstate structure has the least coherence and entanglement over the additive and multiplicative components of the gTPS \cref{structure_H}. Using the duality between optimizing over unitary families of Hamiltonians and algebras, searching over all possible NRC Hamiltonians is equivalent to searching over the equivalence class of isomorphic algebras, so the conjecture can be stated as:

\begin{conj}\hypertarget{conj}{}
    Let $H$ be an NRC Hamiltonian and $\mathcal{A}_f$ be the equivalence class of isomorphic algebras of the form $\mathcal{A}\cong \oplus_{J=1}^{d_\mathcal{Z}} \mathds{1}_{n_J} \otimes \mathcal{L}(\mathbb{C}^{d_J} )$. Then, the algebra corresponding to a gTPS (\cref{structure_H}) for which the Hamiltonian eigenstates have the form $\{\ket{\psi_J}\otimes\ket{\phi_J} \, \vert \, J=1,\dots d_\mathcal{Z}; \, \psi_J = 1,\dots, n_J; \, \phi_J = 1,\dots , d_J \}$ minimizes the $\mathcal{A}$-OTOC LTA over the family of isomorphic algebras. The minimum value is
    \begin{equation}
    \overline{G_{\mathcal{A}}}^{NRC}_{min}=1-\frac{1}{d} \left( \sum_J d_J + \sum_J n_J - d_\mathcal{Z} \right).
\end{equation} 
\end{conj}
\begin{exmp} [\textbf{Quantum reference frames}]
\label{QRF_example}
\par Let us consider the possible quantum reference frame relativity of the $\mathcal{A}$-OTOC LTA minimization. Before applying it to our framework, we provide a brief overview of the perspective-neutral approach to quantum reference frames (QRFs) \cite{de_la_hamette_perspective-neutral_2021,ali_ahmad_quantum_2022,hoehn_quantum_2023}. Intuitively, using a quantum subsystem as a reference frame corresponds to describing the state of the rest of the system in relation to the state of the quantum frame. The reference frames we consider are \textit{internal} to the system, meaning they correspond to quantum subsystems of the full Hilbert space $\mathcal{H}_{tot}$. For simplicity, we shall focus on a full space consisting of two internal QRFs ($R_1$ and $R_2$) and a system of interest $S$, so that $\mathcal{H}_{tot}\cong \mathcal{H}_1 \otimes \mathcal{H}_2 \otimes \mathcal{H}_{S}$. 
\par The basic ingredient of the formalism is a group $\mathcal{G}$, taken to be Abelian here, that constitutes the space of frame orientations, in a similar manner that the Lorentz group SO\textsuperscript{+}(3,1) is the space of classical frame orientations in special relativity \cite{hoehn_quantum_2023,de_la_hamette_perspective-neutral_2021}. For ideal QRFs, the Hilbert spaces $\mathcal{H}_i \cong \mathbb{C}^{\lvert \mathcal{G} \rvert}$ are spanned by the frame configuration states $\ket{g}_i$, $g\in\mathcal{G}$, and furnish a regular unitary representation of $\mathcal{G}$, $U_i^{g^\prime} \ket{g} = \ket{g'g}$. External frame reorientations are given by elements of a unitary tensor product representation of $\mathcal{G}$, $U_{12S}^g\coloneqq U_1^g \otimes U_2^g \otimes U_S^g$. Relational states belong to the physical Hilbert space $\mathcal{H}_{phys}\subset\mathcal{H}_{tot}$, obtained via a coherent group averaging
\begin{equation}
\begin{split}
    &\Pi_{phys}:\mathcal{H}_{tot} \rightarrow \mathcal{H}_{phys}\\
    &\Pi_{phys} \coloneqq \frac{1}{\lvert \mathcal{G} \rvert} \sum_{g\in\mathcal{G}} U_{12S}^g
    \end{split}
\end{equation}
and are invariant under the action of $U_{12S}^g$. $\mathcal{H}_{phys}$ is the gauge-invariant Hilbert space, such that $U_{12S}^g \ket{\psi_{phys}}=\ket{\psi_{phys}} \; \forall \, g\in \mathcal{G},\, \ket{\psi}_{phys} \in \mathcal{H}_{phys}$. Jumping into the perspective of a QRF corresponds to gauge-fixing. For example, we can fix the $R_1$ frame to be in a given orientation $g$. This is achieved via the reduction map
\begin{equation}
\begin{split}
    &\mathcal{R}_{1}^{g}: \mathcal{H}_{phys} \rightarrow \mathcal{H}_2\otimes \mathcal{H}_S\\
    &\mathcal{R}_{1}^{g}=\sqrt{\lvert \mathcal{G} \rvert} \left(\bra{g}_1 \otimes \mathds{1}_2 \otimes \mathds{1}\right) \Pi_{phys}.
\end{split}
\end{equation}
This is a unitary map with the inverse given as \cite[Lemma 21]{hoehn_internal_2022}
\begin{equation}
    (\mathcal{R}_1^g)^{-1} = (\mathcal{R}_1^g)^\dagger = \sqrt{\lvert \mathcal{G} \rvert} \, \Pi_{phys} (\ket{g}_1 \otimes \mathds{1}_2 \otimes \mathds{1}_S).
\end{equation}
Here, $\mathcal{H}_2\otimes \mathcal{H}_S$ is the perspective Hilbert space as ``seen'' from the $R_1$ point of view.
Utilizing the reduction maps, the change from the description of the full system relative to $R_1$ in orientation $g$ to the one relative to $R_2$ in orientation $g^\prime$ is given by
\begin{equation}
    V_{1\rightarrow 2}^{g,g^\prime}=\mathcal{R}_2^{g^\prime} \, (\mathcal{R}_{1}^{g})^\dagger.
\end{equation}
$V_{1\rightarrow 2}^{g,g^\prime}$ is unitary and defines an isomorphism between the perspective Hilbert spaces $\mathcal{H}_2\otimes \mathcal{H}_S$ and $\mathcal{H}_1\otimes \mathcal{H}_S$, while the adjoint action $\hat{V}_{1\rightarrow 2}^{g,g^\prime} (\cdot ) \equiv V_{1\rightarrow 2}^{g,g^\prime} (\cdot ) (V_{1\rightarrow 2}^{g,g^\prime})^\dagger$ is an algebra isomorphism between $\mathcal{L}(\mathcal{H}_2\otimes \mathcal{H}_S)$ and $\mathcal{L}(\mathcal{H}_1\otimes \mathcal{H}_S)$.
\par Let $g_1$ and $g_2$ be the orientations of $R_1$ and $R_2$ and assume that the dynamics from the perspective of $R_1$ are given by a Hamiltonian $H_{\overline{1}}\in \mathcal{L}(\mathcal{H}_2 \otimes \mathcal{H}_S)$. The associated dynamics from the perspective of $R_2$ are, then, given by the Hamiltonian $H_{\overline{2}}=\hat{V}_{1\rightarrow 2}^{g_1,g_2}(H_{\overline{1}})\in \mathcal{L}(\mathcal{H}_1 \otimes \mathcal{H}_S)$. We distinguish between two cases of information scrambling minimization based on the set of algebras of interest:
\par \textit{Perspective-neutral minimization}- Consider a set of algebras of relational observables expressed as $\mathscr{A}_{\overline{1}}=\{\mathcal{A}_{2S}^\mu \subset \mathcal{L} (\mathcal{H}_2 \otimes \mathcal{H}_S)\}_{\mu}$ in the $R_1$-frame and $\mathscr{A}_{\overline{2}}=\{\hat{V}_{1\rightarrow 2}^{g_1,g_2}(\mathcal{A}_{2S}^\mu) \subset \mathcal{L} (\mathcal{H}_1 \otimes \mathcal{H}_S)\}_\mu$ in the $R_2$-frame. The minimization of the $\mathcal{A}$-OTOC LTA can then be performed equivalently either over $\mathscr{A}_{\overline{1}}$ with dynamics given by $H_{\overline{1}}$ or over $\mathscr{A}_{\overline{2}}$ and dynamics given by $H_{\overline{2}}$; in this sense, it is perspective-neutral. The algebras obtained in this way are simply related by $\hat{V}_{1\rightarrow 2}^{g_1,g_2}$, but will, in general, differ in terms of their locality structure with respect to the corresponding perspective Hilbert spaces $\mathcal{H}_2\otimes \mathcal{H}_S$ and $\mathcal{H}_1\otimes \mathcal{H}_S$ \cite{hoehn_quantum_2023}.
\par \textit{Perspectival minimization}- Let $\{\mathcal{A}_{S}^{\mu} \subset \mathcal{L}(\mathcal{H}_S)\}_{\mu}$ be a set of observable algebras of interest in the system $S$. From the perspective of $R_1$, these degrees of freedom are described by $\mathscr{A}_{\overline{1}}=\{\mathds{1}_2 \otimes \mathcal{A}_S^\mu \}_\mu$, while from the perspective of $R_2$ by $\mathscr{A}_{\overline{2}}=\{\mathds{1}_1 \otimes \mathcal{A}_S^\mu \}_\mu$. Although we use the same notation as before for the sets of algebras $\mathscr{A}_{\overline{1}}$, $\mathscr{A}_{\overline{2}}$, we note that they are no longer simply related by $\hat{V}_{1\rightarrow 2}^{g_1,g_2}$; they are now distinct and the $\mathcal{A}$-OTOC LTA minimization will depend on the ``chosen'' perspective.
\par To more concretely illustrate the above observations, let us consider a toy example where the frames $R_1$ and $R_2$ and the system $S$ are qubits. In this case, the frame configuration group is simply the cyclic group of order $2$, $\mathcal{G} = \mathbb{Z}_2=\{ e,g \, \vert \, e^2=e,\, eg=g,\, g^2=e \}$. Also, $\mathcal{H}_i \cong \mathbb{C}^2$ and we identify the configuration states with the computational basis of the $R_1$, $R_2$ qubits, namely $\ket{e}_i\coloneqq \ket{0}_i, \, \ket{g}_i\coloneqq \ket{1}_i$ and, thus, $U_i^{e}=\mathds{1}_i$, $U_i^{g}=\sigma_i^x$, where we will denote as $\sigma_\alpha^{x,y,z}$ the Pauli operators acting on the $\alpha$ subsystem. For simplicity, let the group action on $S$ be the same, namely $U_S^e=\mathds{1}_S$, $U_S^g = \sigma_S^x$. Assuming that both QRFs are in the $e$ orientation, we have $V_{1\rightarrow 2}^{e,e}=\ket{0}_1\otimes \bra{0}_2 \otimes \mathds{1}_S + \ket{1}_1 \otimes \bra{1}_2 \otimes \sigma_S^x$, and $V_{2\rightarrow 1}^{e,e}=\bra{0}_1\otimes \ket{0}_2 \otimes \mathds{1}_S + \bra{1}_1 \otimes \ket{1}_2 \otimes \sigma_S^x$.
\par Let the dynamics in the $R_1$ perspective be given by the Hamiltonian
\begin{equation} \label{Ham_R1}
    H_{\overline{1}}=J_z \, \sigma_2^z \otimes \sigma_S^z + J_x \, \sigma_2^x \otimes \sigma_S^x + J_y \, \sigma_2^y \otimes \sigma_S^y
\end{equation}
where $J_x,J_y,J_z$ are coupling constants. In $R_2$'s perspective, the Hamiltonian is
\begin{equation} \label{Ham_R2}
\begin{split}
    H_{\overline{2}}&=\hat{V}_{1\rightarrow2}^{e,e} (H_{\overline{1}} )=\\
    &= J_z \, \mathds{1}_1 \otimes \sigma_S^z + J_x \, \sigma_1^x \otimes \mathds{1}_S - J_y \, \sigma_1^x \otimes \sigma_S^z
\end{split}
\end{equation}
Using this setup we will now showcase examples for the two types of minimization we conceptually described above.
\begin{enumerate}
    \item As an example of \textit{perspective-neutral} optimization, we consider the so-called natural bipartitions \cite{hoehn_quantum_2023}. Intuitively, these bipartitions correspond to the way the QRF observers divide the ``rest'' of the system into ``other frame'' and $S$ subsystems. Expressing these bipartitions in the $R_1$ frame, we have
    \begin{equation}
    \begin{split}
        &\mathscr{A}_{\overline{1}}=\{\mathcal{A}^1, \mathcal{A}^2 \},\text{ where}\\
        &\mathcal{A}^1 = \mathds{1}_2 \otimes \mathcal{L}(\mathcal{H}_S)\\
        &\hspace{13pt}=\langle \mathds{1}_{2S},\, \mathds{1}_2\otimes \sigma_S^z, \, \mathds{1}_2 \otimes \sigma_S^x , \, \mathds{1}_2 \otimes \sigma_S^y \rangle\\
        &\mathcal{A}^2=\hat{V}_{2\rightarrow 1}^{e,e} (\mathds{1}_1 \otimes \mathcal{L}(\mathcal{H}_S)) = \\
        &\hspace{13pt} =\langle \mathds{1}_{2S}, \, \sigma_2^z \otimes \sigma_S^z , \, \mathds{1}_2 \otimes \sigma_S^x , \, \sigma_2^z \otimes \sigma_S^y \rangle
    \end{split}
    \end{equation}
For $J_x,\, J_y, \, J_z$ all different to each other, the Hamiltonian \cref{Ham_R1} satisfies the NRC and the eigenstates are given by the Bell states $\ket{\phi^{\pm}}=1/\sqrt{2} \left( \ket{0}_2 \otimes \ket{0}_S \pm \ket{1}_2 \otimes \ket{1}_S \right), \, \ket{\psi^\pm}=1\sqrt{2} \left( \ket{0}_2 \otimes \ket{1}_S \pm \ket{1}_2 \otimes \ket{0}_S \right)$. Denoting the Bell states as $\ket{\chi^\lambda}, \, \chi = \phi,\,\psi,\;\lambda=+,\,-$, notice that $\mathcal{A}^2$ is the algebra of observables that acts non-trivially only on $\chi$ and thus corresponds to a virtual bipartition where the Bell states are product states. Then, due to Proposition \ref{nrc_thm},  $\overline{G_{\mathcal{A}^2}(\mathcal{U}_{\overline{1},t})}^t=1/4$, where $\mathcal{U}_{\overline{\alpha},t} (\cdot )=\exp{it \, H_{\overline{\alpha}}} (\cdot ) \exp{-it \, H_{\overline{\alpha}}}$, $\alpha=1,2$, and $\mathcal{A}^2$ minimizes the $\mathcal{A}$-OTOC LTA. Also, $\overline{G_{\mathcal{A}^1}(\mathcal{U}_{\overline{1},t})}^t=3/4$, which saturates the upper-bound $\max\{1-1/d(\mathcal{A} ) , \; 1-1/d(\mathcal{A}^\prime )\}$ of the $\mathcal{A}$-OTOC \cite{andreadakis_scrambling_2023}. Therefore, $\mathcal{A}^2$ is the emergent bipartition, corresponding to $R_1$ and $S$ being distinguishable systems from the perspective of $R_2$, and $R_2$ and $S$ to be ``entangled'' from the perspective of $R_1$. In this sense, $R_2$ remains ``hidden'' from $R_1$, while $R_1$ is ``visible'' to $R_2$.
\item As an example of \textit{perspectival} minimization, consider a family of subalgebras of $\mathcal{A}_S \equiv \{\langle \mathds{1}_{S} , \sigma_S^{\vec{\eta}} \rangle\}_{\vec{\eta}}$, parametrized by the unit vector $\vec{\eta}=\{(\eta_x , \eta_y , \eta_z) \in \mathbb{R}^3 \, \vert \, \eta_x^2 + \eta_y^2 + \eta_z^2 = 1 \}$, where $\sigma_S^{\vec{\eta}}= \eta_x \, \sigma_S^x + \eta_y \, \sigma_S^y + \eta_z \, \sigma_S^z $. These observables correspond to the spin of $S$ in the $\vec{\eta}$ direction and relate to the distinct families of relational observable algebras $\mathscr{A}_1=\{\mathcal{A}_1^{\vec{\eta}} \equiv \langle \mathds{1}_{2S}, \mathds{1}_2 \otimes \sigma_S^{\vec{\eta}} \rangle \}_{\vec{\eta}}$ and $\mathscr{A}_2=\{\mathcal{A}_2^{\vec{\eta}} \equiv \langle \mathds{1}_{1S}, \mathds{1}_1 \otimes \sigma_S^{\vec{\eta}} \rangle \}_{\vec{\eta}}$. In $R_1$'s perspective the dynamics are given by \cref{Ham_R1}, while in $R_2$'s perspective by \cref{Ham_R2}, and the $\mathcal{A}$-OTOC LTA for $\mathcal{A}_1^{\vec{\eta}}$, $\mathcal{A}_2^{\vec{\eta}}$ is given respectively as a function of $\vec{\eta}$ as
\begin{align}
        &\overline{G_{\mathcal{A}_1^{\vec{\eta}}}(\mathcal{U}_{\overline{1},t})}^t=\frac{1}{2} - \frac{\eta_x^4 + \eta_y^4 + \eta_z^4}{8} \label{lta_R1}\\
        &\overline{G_{\mathcal{A}_2^{\vec{\eta}}}(\mathcal{U}_{\overline{2},t})}^t=(1-\eta_z^2) \, \frac{(3+5 \eta_z^2)}{8} \label{lta_R2}
\end{align}
The minimum value of \cref{lta_R1} is $3/8$ and is achieved when any of $\eta_x, \eta_y, \eta_z$ is equal to $\pm1$; from the $R_1$ perspective all three algebras that correspond to the spin directions $\hat{x},\hat{y},\hat{z}$ of $S$ indistinguishably minimize the LTA. The equivalence of the $\hat{x},\hat{y},\hat{z}$ directions is readily anticipated by the fact that the eigenstates of \cref{Ham_R1} are invariant under any $\pi/2$ rotation around any coordinate axis $x,y,z$.   
The minimum value of \cref{lta_R2} is $0$ and is achieved when $\eta_z=\pm1$; from the $R_2$ perspective, the algebra corresponding to the spin direction $\hat{z}$ of $S$ is dynamically preferred, maximally retaining its informational content. This is intuitive since the spin of $S$ in the $z$-direction is a conserved quantity of the Hamiltonian in \cref{Ham_R2}. This shows that using the $\mathcal{A}$-OTOC minimization criterion, the dynamically preferred partition of the system $S$ differs for observers in different QRFs.
\end{enumerate}
\end{exmp}


\begin{exmp}[\textbf{Many-body Emergent Bipartitions}]

\begin{figure*}
\centering
\begin{subfigure}{.5\textwidth}
  \centering
  \includegraphics[width=1\linewidth]{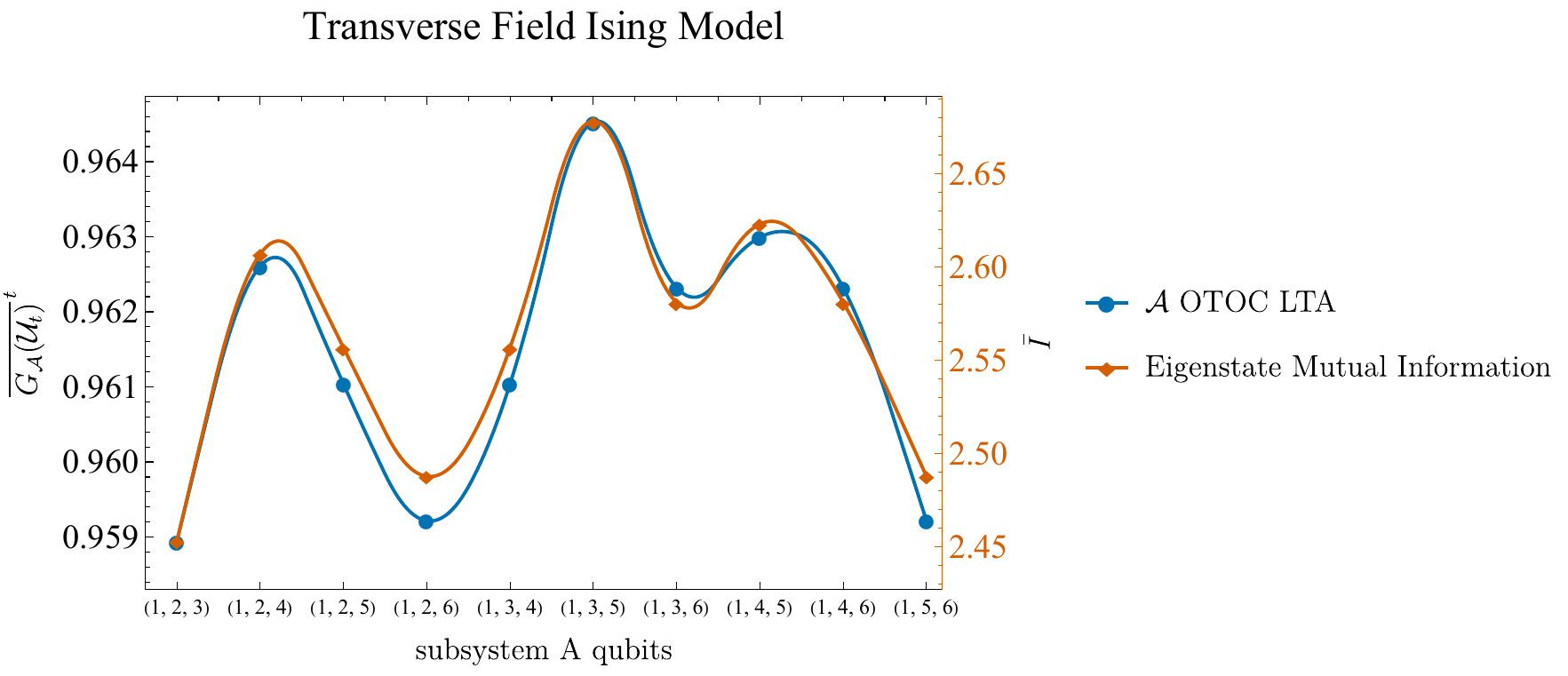}
 \caption{}
\end{subfigure}%
\begin{subfigure}{.5\textwidth}
  \centering
  \includegraphics[width=1\linewidth]{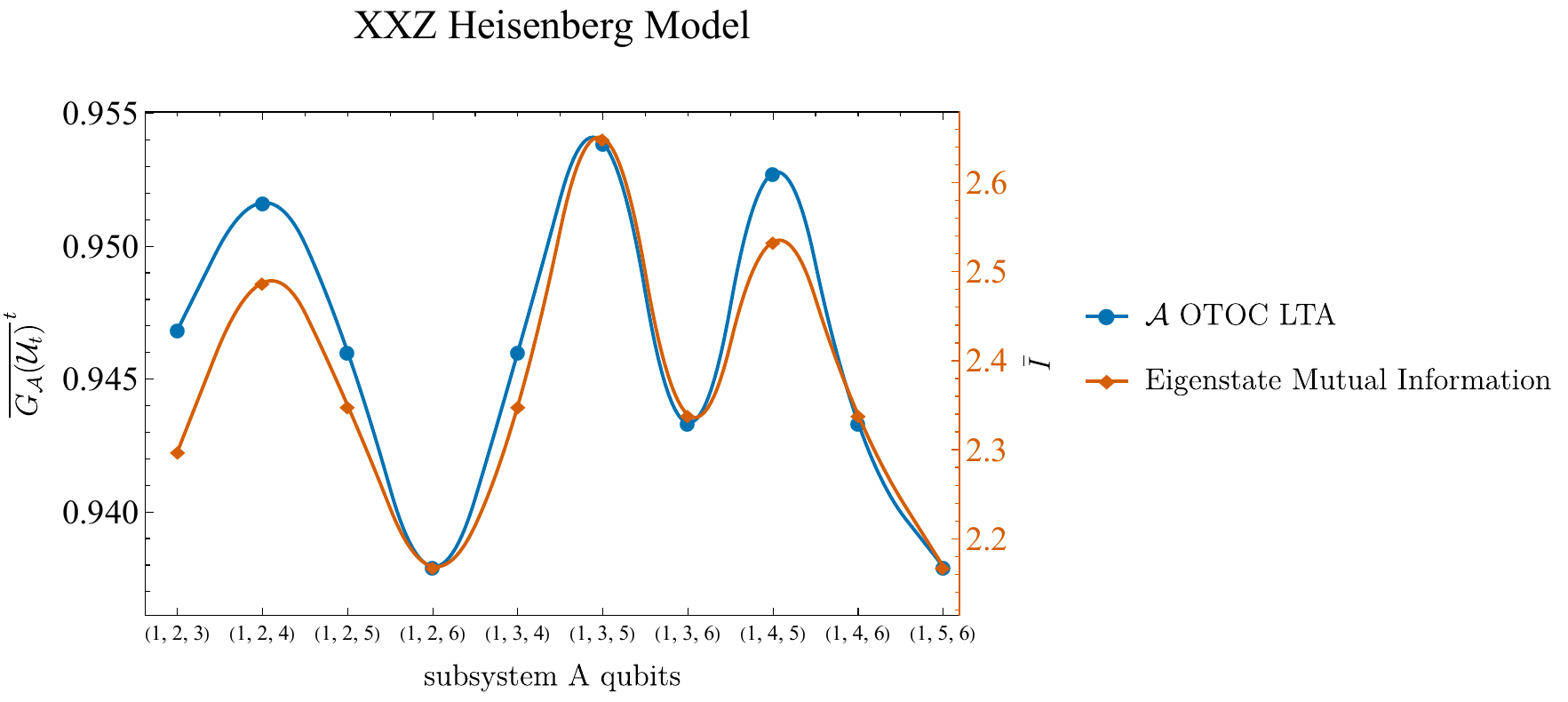}
  \caption{}
\end{subfigure}
\caption{The dependence of the bipartite $\mathcal{A}$-OTOC LTA $\overline{G_{\mathcal{A}}(\mathcal{U}_t)}^t$ and the average eigenstate mutual information $\bar{I}$ on the choice of bipartition for (a) the transverse field Ising model (TFIM) and (b) the XXZ Heisenberg model with $N=6$ qubits. The qualitative behavior of $\overline{G_{\mathcal{A}}(\mathcal{U}_t)}^t$ and $\bar{I}$ is identical, showing that, for the chosen many-body models, the bipartite $\mathcal{A}$-OTOC LTA probes the amount of correlations of the Hamiltonian eigenstates across the bipartition A:B.}

\label{fig_mutual}
\end{figure*}
    
\par We now focus on the important, special case of a bipartite algebra $\mathcal{A}$, with $\mathcal{H}\cong \mathcal{H}_A \otimes \mathcal{H}_B$ and $\mathcal{A}\cong \mathds{1}_A \otimes \mathcal{L}(\mathcal{H}_B)$. Here, $G_{\mathcal{A}}(\mathcal{U}_t)$ coincides with the operator entanglement of $U_t$ across the A:B bipartition \cite{styliaris_information_2021}, where $\mathcal{U}_t (\cdot )= U_t (\cdot ) U_t^\dagger$.
\par For a bipartite algebra, the Gaussian scrambling rate \cref{gaus} that dictates the short-time behavior of the $\mathcal{A}$-OTOC is \cite{zanardi_operational_2023}
\begin{equation} \label{short_bip}
    \tau_S^{-1}= \frac{1}{\sqrt{d}}\left\lVert{H-\frac{\mathds{1}_A}{d_A} \otimes \Tr_A(H) -\Tr_B(H)\otimes\frac{\mathds{1}_B}{d_B}}\right\rVert_2,
\end{equation}
assuming for simplicity that $Tr(H)=0$. \cref{short_bip} shows that the short-time behavior of the bipartite $\mathcal{A}$-OTOC depends exactly on the strength of the interaction part of the Hamiltonian between A and B.
\par For the bipartite $\mathcal{A}$-OTOC NRC LTA of \cref{NRC} we see that \cite{styliaris_information_2021}
\begin{equation} \label{long_bip}
\begin{split}
&d^2\left(1-\overline{G (\mathcal{U}_t )}^{NRC}\right) =\\
&\hspace{5pt}=\sum_{X = \{ A,B \}} \left(\sum_{k,l=1}^d \langle\rho_k^{X},\rho_l^{X} \rangle^2 - \frac{1}{2}\sum_{k=1}^d \langle\rho_k^{X},\rho_k^{X} \rangle^2\right)
\end{split}
\end{equation}
where $R^{(X)}_{kl} \coloneqq \langle\rho_k^{X},\rho_l^{X} \rangle$ is the Gram matrix of the reduced Hamiltonian eigenstates on the $X$ subsystem. \cref{long_bip} shows that the bipartite $\mathcal{A}$-OTOC LTA is intimately related to the entanglement structure of the Hamiltonian eigenstates that depends on the full details of the Hamiltonian operator, well beyond simply the interaction strength.

\par In what follows, we consider a background tensor product structure (TPS) $\mathcal{H} \cong \bigotimes_i \mathcal{H}_i$ and compare the short- and long-time behavior of the $\mathcal{A}$-OTOC for algebras that act non-trivially on a subset of the subsystems $\mathcal{H}_i$. Specifically, we consider a spin-chain of $N$ qubits, $\mathcal{H}\cong \left(\mathbb{C}^2\right)^N$ with open boundary conditions and dynamics given by: i) the XXZ Heisenberg model $H_{XXZ}=\sum_{i=1}^{N} \left(J_X \left( \sigma_i^x \sigma_{i+1}^x + \sigma_i^y \sigma_{i+1}^y \right) + J \sigma_i^z \sigma_{i+1}^z \right)$ with $J_X=-0.4, \, J=-1$, ii) the transverse field Ising model (TFIM) $H_I = \sum_{i=1}^N \left( h \sigma_{i}^z + g \sigma_i^x \right) -\sum_{i=1}^{N-1} \sigma_i^z \sigma_{i+1}^z$ with $h=-0.5, \, g=1.05$. Let $N$ be even and $\mathscr{A}$ be the set of algebras that act non-trivially on exactly $N/2$ (not necessarily contiguous) qubits. We are interested in the emergent spatial bipartition for the short- and long-time limits based on the \hyperlink{principle_1}{principle} of minimal scrambling. Clearly, the bipartition that minimizes the interaction Hamiltonian, and thus the Gaussian scrambling rate, for either of the models is the contiguous one: $\mathcal{H}_A = \bigotimes_{i=1}^{N/2} \mathcal{H}_i$, $\mathcal{H}_B = \bigotimes_{i=N/2+1}^{N} \mathcal{H}_i$. 
\par Note that for the above coupling constants, the XXZ model satisfies NRC\textsuperscript{+} as there are no gap degeneracies, while the Ising model is chaotic, satisfying NRC \cite{styliaris_information_2021}. Using \cref{NRCp} for the XXZ model and \cref{NRC} for the Ising model, we calculate for $N=6$ the $\mathcal{A}$-OTOC LTA for all possible choices of algebras in $\mathscr{A}$. We find that the bipartition that minimizes the $\mathcal{A}$-OTOC LTA coincides with the short-time one, $\mathcal{H}_A= \mathcal{H}_1 \otimes \mathcal{H}_2 \otimes \mathcal{H}_3$, for the Ising model, while for the XXZ model there are two non-contiguous long-time preferred bipartitions, $\mathcal{H}_A= \mathcal{H}_1 \otimes \mathcal{H}_2 \otimes \mathcal{H}_6$ and $\mathcal{H}_A= \mathcal{H}_1 \otimes \mathcal{H}_5 \otimes \mathcal{H}_6$.
\par As is clear from \cref{long_bip}, the $\mathcal{A}$-OTOC LTA is related to the eigenstate correlations between either subsystem of a bipartition. In order to make this relation more concrete, we compute for the models above an average eigenstate mutual information
\begin{equation}
    \bar{I}\equiv\frac{1}{M} \sum_{i=1}^M I\left( \rho_{E_i} \right)
\end{equation}
where $\rho_{E_i} \coloneqq \Pi_i / \Tr( \Pi_i )$ is the uniform pure state ensemble of eigenstates in a given energy $E_i$ and $I(\rho ) \coloneqq S(\rho^A) + S(\rho^B) - S(\rho)$ is the mutual information of $\rho$ across the $A:B$ bipartition. As shown in \cref{fig_mutual}, the qualitative behavior of the $\mathcal{A}$-OTOC as a function of the chosen bipartition is identical to that of $\bar{I}$. Quite intuitively, for the many-body systems considered here, the information scrambling between $A$ and $B$ in long times is controlled by the amount of eigenstate mutual information, and its minimization over $\mathscr{A}$ is related to the subsystem emergence in the long-time limit.
\end{exmp}

\subsection{Numerical Evidence for Conjecture} \label{num_evidence}

We make use of the algebra-Hamiltonian duality of the NRC LTA by fixing an algebra $\mathcal{A}$ and searching over the space of Hamiltonians for a violation of the conjecture. As seen in \cref{NRC}, the NRC LTA depends only on the Hamiltonian eigenbasis, which can be represented by a unitary matrix with eigenstates as the columns. Given an underlying basis, this representation is unique up to permutations of the columns. Succinctly, for Hamiltonian eigenbasis (unitary) $E$ and algebra $\mathcal{A}$, the LTA is a function $f_{\mathcal{A}}(E)$.

\par
For any arbitrary unitary $U$, $UE$ is also a unitary corresponding to the eigenbasis of some other class of Hamiltonians, and for any eigenbasis $Q$, there exists a $U$ such that $Q = UE$. Fixing $E = I$ for convenience, which corresponds to a Hamiltonian eigenbasis satisfying the conjecture\footref{distinguished_basis}, the LTA is a function $f_{\mathcal{A}}(U)$\footnote{It is important to recognize that $U$ here is distinct from the time-evolution unitary $U_t$ in previous formulae.}. We can now search over all unitaries to try to find a $U$ such that $f_{\mathcal{A}}(U)$ is less than the conjectured minimum. To do this, for a fixed algebra, we use an algorithm devised by Abrudan et al. \cite{abrudan_steepest_2008} to perform gradient descent of the LTA on the space of unitary matrices (see \cref{grad_appendix} for elaboration). The algorithm is run until convergence, beginning with a random unitary. At this point, we compare the observed LTA value with the conjectured minimum.

\par
 To systematically search for a violation of the conjecture, we need to perform this gradient descent for all algebras of a given Hilbert space dimension.
 Recall that each algebra has a structure given by \cref{structure_A}. This allows us to characterize each algebra by $d_Z$ and the set of pairs $ \{(n_J, d_J) \}_{J=1}^{d_Z}$. Note that these values correspond to an equivalence class of algebras that are identical up to intra-sector unitary conjugation, which can be thought of as just performing a change of basis within a sector. Because we are already performing a search over unitaries, it is sufficient to select just one representative algebra from each equivalence class on which to run the algorithm.

 \par
 To generate these algebras, given a Hilbert space dimension $d$, we generate all unique sets of positive integers which sum to $d$. Each integer corresponds to the dimension of a block in the direct sum \cref{structure_A}, and $d_Z$ is the order of the set. The ordering of elements in a given set does not matter, since the rearrangement of blocks in the direct sum does not affect the algebra structure. Next, we decompose each integer into a product of two factors, corresponding to $n_J$ and $d_J$; the full set of pairs corresponds to an equivalence class of algebras. We continue to perform this decomposition until we generate all unique sets of pairs, i.e., every equivalence class of algebras. Finally, we run the unitary gradient descent algorithm on each algebra class.
 \par
  We were able to generate all algebra classes for Hilbert spaces up to $d = 40$. 
 For each class generated, the resultant LTA value \textit{equals} the conjectured minimum. This provides some numerical evidence that the conjectured minimum is at least a local minimum for small systems.

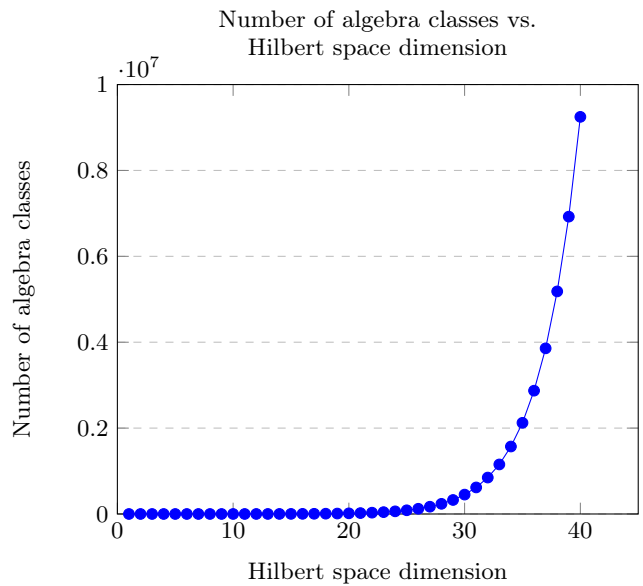
\begin{figure}
   \begin{tikzpicture}
\begin{axis}[align=center,
    title={Number of algebra classes vs. \\ Hilbert space dimension},
    xlabel={Hilbert space dimension},
    ylabel={Number of algebra classes},
    xmin=0, xmax=45,
    ymin=0, ymax=10000000,
    xtick={0,10,20,30,40},
    ytick={0,2000000,4000000,6000000,8000000,10000000},
    legend pos=north west,
    ymajorgrids=true,
    grid style=dashed,
]

\addplot[
    color=blue,
    mark = *
    ]
    coordinates {(1, 1)(2, 3)(3, 5)(4, 11)(5, 17)(6, 34)(7, 52)(8, 94)(9,145)(10, 244)(11, 370)(12, 603)(13, 899)(14, 1410)(15, 2087)(16, 3186)(17, 4650)(18, 6959)(19, 10040)(20, 14750)(21, 21077)(22, 30479)(23, 43120)(24, 61574)(25, 86308)(26, 121785)(27, 169336)(28, 236475)(29, 326201)(30, 451402)(31, 618135)(32, 848209)(33, 1153733)(34, 1571063)(35, 2123325)(36, 2871419)(37, 3857569)(38, 5182999)(39, 6924303)(40, 9247955)
    };
    
\end{axis}
\end{tikzpicture}

\caption{The number of algebra classes scales exponentially in Hilbert space dimension, making higher dimensional tests of the conjecture difficult.}

\end{figure}

\section{Conclusion}\label{conclusion}
\par {We have studied the long-time properties of quantum information scrambling of algebras of observables $\mathcal{A}$, quantified by the long-time average (LTA) of the $\mathcal{A}$-OTOC. In systems with energy spectra that satisfy non-resonant conditions, we have derived simplified expressions for the $\mathcal{A}$-OTOC LTA in terms of the Hamiltonian eigenprojectors and projections onto the algebras $\mathcal{A}$ and $\mathcal{A}^\prime$. In the presence of resonances, we have shown how these non-resonant approximations can be used as a proxy for the behavior of the exact LTA for a unitary family of stabilizer algebras as well as the dynamics generated by a Heisenberg Hamiltonian. We expect that such a procedure will be of practical significance in determining the long-time scrambling properties of algebras of observables in physical systems of interest, as computing the non-resonant approximations is considerably less computationally complex than calculating the exact LTA value.}
\par {Based on the above results, we have} extended the use of the $\mathcal{A}$-OTOC as a criterion for quantum mereology into the ``macroscopic'' regime. We propose that, given a system Hamiltonian, the minimization of the $\mathcal{A}$-OTOC LTA selects an algebra of observables which corresponds to a preferred Hilbert space partition that best retains its informational structure under the dynamics. {Using the analytic formula for the LTA and a duality in optimizing the $\mathcal{A}$-OTOC over unitary families of algebras and unitary families of Hamiltonians, we have performed analytic minimizations of the $\mathcal{A}$-OTOC LTA. For certain unitary families of NRC Hamiltonians, including the family of all NRC Hamiltonians in a bipartite Hilbert space, we have shown that the $\mathcal{A}$-OTOC LTA is minimized when the eigenstates are unentangled across the virtual bipartitions induced by $\mathcal{A}$. Based on these analytic results and some numerical evidence, we conjecture that the minimization of the $\mathcal{A}$-OTOC LTA over all isomorphic algebras with fixed structural dimensions $d_\mathcal{Z},\{d_J\},\{n_J\}$, is achieved by algebras that induce a gTPS (\cref{structure_H}) for which the eigenstates belong to a unique superselection sector and are unentangled across the virtual bipartitions.}

\par As a further application, we illustrated via a toy example how our $\mathcal{A}$-OTOC LTA minimization framework depends on the choice of an internal quantum reference frame (QRF) in the context of the perspective-neutral approach to QRFs. Finally, we numerically studied the $\mathcal{A}$-OTOC LTA in the case of bipartite algebras for certain quantum many-body models and showed that its behavior with respect to the chosen bipartition is connected to the average eigenstate mutual information between the corresponding subsystems.
\par The work here can be extended in several paths. Further analytical and numerical studies are needed to investigate the validity of the conjecture for the algebra that minimizes the $\mathcal{A}$-OTOC for NRC Hamiltonian dynamics. It is also natural to consider the long-time behavior of the $\mathcal{A}$-OTOC in open quantum systems, where there is a competition between the entropic contributions of information scrambling and decoherence \cite{zanardi_information_2021,andreadakis_scrambling_2023}.

\section{Acknowledgments}
 FA and PZ acknowledge partial support from the NSF award PHY-2310227. FA acknowledges partial support from the Gerondelis foundation.
 This research was (partially) sponsored by the Army Research Office and 
was accomplished under Grant Number W911NF-20-1-0075.  
The views and conclusions contained in this document are those of 
the authors and should not be interpreted as representing the official 
policies, either expressed or implied, of the Army Research Office
 or the U.S. Government. The U.S. Government is authorized to reproduce and 
distribute reprints for Government purposes notwithstanding any copyright
 notation herein.
\bibliographystyle{apsrev4-2}
\bibliography{Mereology}

\onecolumngrid
\newpage
\appendix*
\section{}\label{appendix}

\subsection{Proof of \cref{prop_NRC}}
Let $\mathcal{H} \cong \mathbb{C}^d$ be a Hilbert space. For a Hamiltonian evolution $U_t=\exp(itH)$ the $\mathcal{A}$-OTOC can be expressed as \cite{andreadakis_scrambling_2023}:
\begin{equation} \label{a-otoc}
G_\mathcal{A} (\mathcal{U}_t)= 1-\frac{1}{d} \Tr( S \, \Omega_\mathcal{A} \, \mathcal{U}_t^{\otimes 2} \left( \Omega_{\mathcal{A}^\prime} \right))
\end{equation}
where $\mathcal{U}_t (\cdot )= U_t (\cdot ) U_t^\dagger$, $\Omega_\mathcal{A} = \sum_\alpha e_\alpha \otimes e_\alpha^\dagger$, $\Omega_{\mathcal{A}^\prime}= \sum_\gamma f_\gamma \otimes f_\gamma^\dagger$ and $\{e_\alpha \}_{\alpha=1}^{d(\mathcal{A})}$, $\{f_\gamma \}_{\gamma=1}^{d(\mathcal{A}^\prime)}$ are appropriate orthonormal bases of $\mathcal{A}$ and $\mathcal{A}^\prime$ repsectively. We can choose these algebra bases in terms of the decomposition \cref{structure_H} as
\begin{equation} \label{bases_formula}
    \begin{split}
        &f_\gamma = \ketbra{p_J}{q_J} \otimes \frac{\mathds{1}_{d_J}}{\sqrt{n_J}}, \, \gamma=(J,p_J,q_J); \, J=1,\dots , d_\mathcal{Z}; \, p_J,q_J = 1, \dots , n_J \\
        &e_\alpha=\frac{\mathds{1}_{n_J}}{\sqrt{d_J}} \otimes \ketbra{k_J}{l_J},\, \alpha=(J,k_J,l_J); \, J=1,\dots , d_\mathcal{Z}; \, k_J,l_J = 1, \dots , d_J
    \end{split}
\end{equation}
where we used a basis $\mathbb{B}=\{ \ket{p_J} \otimes \ket{k_J}\, \vert \, J=1,\dots , d_\mathcal{Z}; \, p_J = 1,\dots , n_J ; \, k_J=1,\dots , d_J \}$ of $\mathcal{H}$. Using the spectral decomposition $H=\sum_{k=1}^M E_k \Pi_k$, we have
\begin{equation}
    \begin{split}
        \overline{\mathcal{U}_t^{\otimes 2}(\cdot )}^t&=\lim_{T\rightarrow \infty} \frac{1}{T} \int_0^T \sum_{k,l,m,n=1}^M\exp(it (E_k+E_l-E_m-E_n)) \Pi_k \otimes \Pi_l (\cdot ) \Pi_m \otimes \Pi_n \; dt \\
        &= \sum_{k,l,m,n=1}^M \delta_{E_k+E_l,E_m+E_n} \, \Pi_k \otimes \Pi_l (\cdot ) \Pi_m \otimes \Pi_n 
    \end{split}
\end{equation}
Substituting the explicit expressions on \cref{a-otoc} and using the identity
\begin{equation}
    \Tr(S \, A \otimes B ) = \Tr(AB)
\end{equation}
we have
\begin{equation} \label{proof_exact}
\begin{split}
    \overline{G_\mathcal{A} (\mathcal{U}_t)}^t&=1-\frac{1}{d}  \sum_{k,l,m,n=1}^M \delta_{E_k+E_l,E_m+E_n} \Tr(\sum_{\alpha=1}^{d(\mathcal{A})} \sum_{\gamma=1}^{d(\mathcal{A}^\prime)} e_\alpha \Pi_k f_\gamma \Pi_m e_\alpha^\dagger \Pi_l f_\gamma^\dagger \Pi_n)\\
    &= 1-\frac{1}{d}  \sum_{k,l,m,n=1}^M \delta_{E_k+E_l,E_m+E_n} \sum_{\gamma=1}^{d(\mathcal{A}^\prime)} \langle \mathbb{P}_{\mathcal{A}^\prime}(\Pi_k f_\gamma \Pi_m), \mathbb{P}_{\mathcal{A}^\prime}(\Pi_n f_\gamma \Pi_l)\rangle
\end{split}
\end{equation}
where we used that $\mathbb{P}_{\mathcal{A}^\prime} (\cdot ) = \sum_\alpha e_\alpha (\cdot ) e_\alpha^\dagger$ is a projector.
\par i) The NRC\textsuperscript{+} condition requires that there are no degenerate gaps, which implies that $E_k+E_l=E_m+E_n \Leftrightarrow (k=m, \; l=n) \text{ or }(k=n, \; l=m) \Leftrightarrow \delta_{E_k+E_l, E_m+E_n}= \delta_{k,m} \delta_{l,n} + \delta_{k,n} \delta_{l,m} - \delta_{k,l} \delta_{k,m} \delta_{k,n})$. So, \cref{proof_exact} becomes
\begin{equation} \label{nrcp}
\begin{split}
    \overline{G_\mathcal{A} (\mathcal{U}_t)}^{NRC^+}&=1-\frac{1}{d} \sum_{\alpha=1}^{d(\mathcal{A})} \sum_{\gamma=1}^{d(\mathcal{A}^\prime)} \left(\sum_{k,l=1}^M \left(\Tr(e_\alpha \Pi_k f_\gamma \Pi_k e_\alpha^\dagger \Pi_l f_\gamma^\dagger \Pi_l)+\Tr(e_\alpha \Pi_k f_\gamma \Pi_l e_\alpha^\dagger \Pi_l f_\gamma^\dagger \Pi_k)\right)-\right.\\
    &\hspace{100pt}\left.-\sum_{k=1}^M\Tr(e_\alpha \Pi_k f_\gamma \Pi_k e_\alpha^\dagger \Pi_k f_\gamma^\dagger \Pi_k)\right)=\\
    &=1 - \frac{1}{d}\left[ \sum\limits_{\gamma=1}^{d(\mathcal{A}^\prime)} \left(\lVert \mathbb{P}_{\mathcal{A}^\prime}(D_H (f_{\gamma}))\rVert_2^2 - \frac{1}{2}\sum\limits_{k=1}^M \lVert \mathbb{P}_{\mathcal{A}^\prime}(\Pi_k f_{\gamma} \Pi_k)\rVert_2^2 \right) +\right.\\
    &\hspace{50pt}\left.+ \sum\limits_{\alpha=1}^{d(\mathcal{A})}\left(\lVert \mathbb{P}_{\mathcal{A}}(D_H (e_{\alpha}))\rVert_2^2 - \frac{1}{2}\sum\limits_{k=1}^M \lVert \mathbb{P}_{\mathcal{A}}(\Pi_k e_{\alpha} \Pi_k)\rVert_2^2 \right)\right]
\end{split}
\end{equation}
\par ii) The NRC condition requires in addition that the energy levels are non-degenerate, so the eigenprojectors $\Pi_k$ are 1-dimensional, i.e. $\Pi_k=\ketbra{\phi_k}{\phi_k} \; \forall \, k=1,\dots ,d$. Notice that
\begin{equation} \label{nrc_1}
    \begin{split}
        \sum_{\gamma=1}^{d(\mathcal{A}^\prime)}\left\lVert \mathbb{P}_{\mathcal{A}^\prime}(D_H (f_{\gamma}))\right\rVert_2^2&=\sum_{\gamma=1}^{d(\mathcal{A}^{\prime})} \left\lVert \sum_{k=1}^d \mel{\phi_k}{f_\gamma}{\phi_k} \mathbb{P}_{\mathcal{A}^\prime} (\ketbra{\phi_k}{\phi_k})\right\rVert_2^2=\\
        &=\sum_{\gamma=1}^{d(\mathcal{A}^\prime)}\sum_{k,l=1}^d \mel{\phi_l}{f_\gamma}{\phi_l}\mel{\phi_k}{f_\gamma^\dagger}{\phi_k} \langle \mathbb{P}_{\mathcal{A}^\prime} (\ketbra{\phi_k}{\phi_k}) , \mathbb{P}_{\mathcal{A}^\prime} ( \ketbra{\phi_l}{\phi_l} ) \rangle = \\
        &=\sum_{k,l=1}^d \langle \ketbra{\phi_k}{\phi_l}, \mathbb{P}_\mathcal{A} (\ketbra{\phi_k}{\phi_l} ) \rangle \, \langle \mathbb{P}_{\mathcal{A}^\prime} (\ketbra{\phi_k}{\phi_k}) , \mathbb{P}_{\mathcal{A}^\prime} ( \ketbra{\phi_l}{\phi_l} ) \rangle = \Tr( R^{(0),\mathcal{A}} \, R^{(1),\mathcal{A}^\prime} )
    \end{split}
\end{equation}
where $R_{lk}^{(0),\mathcal{A}} \coloneqq \left\lVert \mathbb{P}_{\mathcal{A}}\left(\ket{\phi_k}\bra{\phi_l} \right) \right\rVert_2^2$, $R_{kl}^{(1),\mathcal{A}} \coloneqq  \left\langle\mathbb{P}_{\mathcal{A}}\left(\Pi_k\right),  \mathbb{P}_{\mathcal{A}}\left(\Pi_l\right)\right\rangle$. Similarly,
\begin{equation} \label{nrc_2}
    \begin{split}
        &\sum_{\alpha=1}^{d(\mathcal{A})}\lVert \mathbb{P}_{\mathcal{A}}(D_H (e_{\alpha}))\rVert_2^2=\Tr( R^{(0),\mathcal{A}^\prime}\, R^{(1),\mathcal{A}})\\
        &\sum_{k=1}^d\sum_{\gamma=1}^{d(\mathcal{A}^\prime)}\lVert \mathbb{P}_{\mathcal{A}^\prime}(\Pi_k f_{\gamma} \Pi_k)\rVert_2^2=\Tr( R_D^{(0),\mathcal{A}} \, R_D^{(1),\mathcal{A}^\prime} )\\
        &\sum_{k=1}^d\sum_{\alpha=1}^{d(\mathcal{A})}\lVert \mathbb{P}_{\mathcal{A}}(\Pi_k e_{\alpha} \Pi_k)\rVert_2^2= \Tr( R_D^{(0),\mathcal{A}^\prime} \, R_D^{(1),\mathcal{A}} ) 
    \end{split}
\end{equation}
where $R_D=\diag(R)$. Plugging \cref{nrc_1,nrc_2} in \cref{nrcp} one gets \cref{NRC}.

\subsection{Proof of Proposition \ref{nrc_thm}}
Recall that given the Hilbert space decomposition \cref{structure_H}, we can express the projectors over $\mathcal{A}$ and $\mathcal{A}^\prime$ as:

\begin{equation}
\begin{split}
&\mathbb{P}_{\mathcal{A}^\prime} (\cdot ) = \bigoplus_{J=1}^{d_\mathcal{Z}} \left( \Tr_{d_J}(\cdot ) \otimes \frac{\mathds{1}_{d_J}}{d_J} \right)\\
&\mathbb{P}_{\mathcal{A}} (\cdot ) = \bigoplus_{J=1}^{d_\mathcal{Z}} \left(  \frac{\mathds{1}_{n_J}}{n_J} \otimes \Tr_{n_J}(\cdot ) \right)
\end{split}
\end{equation}
Also, we have a partial-trace trick:
\begin{equation}
\Tr_A(\Tr_B (\ket{\phi_k}\bra{\phi_l}) \, \Tr_B( \ket{\phi_l} \ket{\phi_k})) = \Tr_B( \Tr_A (\ket{\phi_k}\bra{\phi_k}) \, \Tr_A( \ket{\phi_l} \ket{\phi_l}))
\end{equation}
Using the above we have
\begin{equation} \label{alt1}
    \begin{split}
        &\Tr( R^{(0),\mathcal{A}} \, R^{(1),\mathcal{A}^\prime} )=\\
        &\hspace{30pt}=\sum_{k,l=1}^d \left\lVert \bigoplus_{J=1}^{d_\mathcal{Z}} \frac{\mathds{1}_{n_J}}{n_J}\otimes \Tr_{n_J}(\ketbra{\phi_k^J}{\phi_l^J}) \right\rVert_2^2 \, \left\langle \bigoplus_{K=1}^{d_\mathcal{Z}} \Tr_{d_K}(\ketbra{\phi_k^K}{\phi_k^K} )\otimes \frac{\mathds{1}_{d_K}}{d_K}, \bigoplus_{L=1}^{d_\mathcal{Z}}\Tr_{d_L}(\ketbra{\phi_l^L}{\phi_l^L}) \otimes \frac{\mathds{1}_{d_L}}{d_L}\right\rangle=\\
        &\hspace{30pt}=\sum_{k,l=1}^d \sum_{J=1}^{d_\mathcal{Z}} \left(\frac{\langle \Tr_{n_J}(\ketbra{\phi_l^J}{\phi_k^J}) , \Tr_{n_J}(\ketbra{\phi_l^J}{\phi_k^J})\rangle}{n_J} \right)\, \sum_{K=1}^{d_\mathcal{Z}} \left(\frac{\langle \Tr_{d_K}(\ketbra{\phi_k^K}{\phi_k^K}),\Tr_{d_K}(\ketbra{\phi_l^K}{\phi_l^K} )\rangle}{d_K} \right)=\\
        &\hspace{30pt}= \sum_{k,l=1}^d \sum_{J=1}^{d_\mathcal{Z}} \left(\frac{\langle\rho_k^{n_J},\rho_l^{n_J} \rangle}{n_J} \right) \, \sum_{K=1}^{d_\mathcal{Z}} \left(\frac{\langle \rho_k^{n_K} , \rho_l^{n_K} \rangle}{d_K} \right)
    \end{split}
\end{equation}
where $\ket{\phi_k^J}$ is the projection of $\ket{\phi_k}$ on the $J^{th}$ sector and $\rho_k^{n_J} \coloneqq \Tr_{d_J} (\ketbra{\phi_k^J}{\phi_k^J})$. Similarly,
\begin{equation} \label{alt2}
    \begin{split}
        &\Tr( R^{(0),\mathcal{A}^\prime}\, R^{(1),\mathcal{A}})= \sum_{k,l=1}^d \sum_{J=1}^{d_\mathcal{Z}} \left( \frac{\langle\rho_k^{d_J},\rho_l^{d_J} \rangle}{d_J} \right) \, \sum_{K=1}^{d_\mathcal{Z}} \left(\frac{\langle \rho_k^{d_K} , \rho_l^{d_K} \rangle}{n_K} \right)\\
        &\Tr( R_D^{(0),\mathcal{A}} \, R_D^{(1),\mathcal{A}^\prime} )= \sum_{k=1}^d \sum_{J=1}^{d_\mathcal{Z}} \left(\frac{\langle\rho_k^{n_J},\rho_k^{n_J} \rangle}{n_J} \right)\, \sum_{K=1}^{d_\mathcal{Z}} \left(\frac{\langle \rho_k^{n_K} , \rho_k^{n_K} \rangle}{d_K} \right)\\
        &\Tr( R_D^{(0),\mathcal{A}^\prime} \, R_D^{(1),\mathcal{A}} )=\sum_{k=1}^d \sum_{J=1}^{d_\mathcal{Z}} \left(\frac{\langle\rho_k^{d_J},\rho_k^{d_J} \rangle}{d_J} \right)\, \sum_{K=1}^{d_\mathcal{Z}} \left(\frac{\langle \rho_k^{d_K} , \rho_k^{d_K} \rangle}{n_K} \right)
    \end{split}
\end{equation}
Using \cref{alt1,alt2} in \cref{NRC} we have
\begin{equation} \label{alt_NRC}
\begin{split}
&d\left(1-\overline{G_\mathcal{A} (\mathcal{U}_t )}^{NRC}\right) =\\
&\hspace{49pt}=\sum_{k,l=1}^d \sum_{J=1}^{d_\mathcal{Z}} \left(\frac{\langle\rho_k^{d_J},\rho_l^{d_J} \rangle}{n_J} \right) \, \sum_{K=1}^{d_\mathcal{Z}} \left(\frac{\langle \rho_k^{d_K} , \rho_l^{d_K} \rangle}{d_K} \right) - \frac{1}{2}\sum_{k=1}^d \sum_{J=1}^{d_\mathcal{Z}} \left(\frac{\langle\rho_k^{d_J},\rho_k^{d_J} \rangle}{n_J} \right)\, \sum_{K=1}^{d_\mathcal{Z}} \left(\frac{\langle \rho_k^{d_K} , \rho_k^{d_K} \rangle}{d_K} \right) +\\
&\hspace{58pt}+ \sum_{k,l=1}^d \sum_{J=1}^{d_\mathcal{Z}} \left( \frac{\langle\rho_k^{n_J},\rho_l^{n_J} \rangle}{d_J} \right) \, \sum_{K=1}^{d_\mathcal{Z}} \left(\frac{\langle \rho_k^{n_K} , \rho_l^{n_K} \rangle}{n_K} \right) - \frac{1}{2} \sum_{k=1}^d \sum_{J=1}^{d_\mathcal{Z}} \left(\frac{\langle\rho_k^{n_J},\rho_k^{n_J} \rangle}{d_J} \right)\, \sum_{K=1}^{d_\mathcal{Z}} \left(\frac{\langle \rho_k^{n_K} , \rho_k^{n_K} \rangle}{n_K} \right)
\end{split}
\end{equation}


\par Let us calculate the $\mathcal{A}$-OTOC NRC LTA when the eigenstates are of the form $\ket{\phi_k}=\ket{J,p} \otimes \ket{J,m}$ where $\ket{J,p} \in \mathbb{C}^{n_J}$ and $\ket{J,m} \in \mathbb{C}^{d_J}$. Using the multi-index notation $k=(R,p,m)$:
\begin{equation}
\begin{split}
&\rho_{R,p,m}^{n_J}=\delta_{J,R} \, \ket{R,p}\bra{R,p}\\
&\langle \rho_{R,p,m}^{n_J} , \rho_{W,s,t}^{n_J} \rangle = \delta_{J,R} \, \delta_{R,W} \, \delta{p,s} \\
&\langle \rho_{R,p,m}^{d_J} , \rho_{W,s,t}^{d_J} \rangle = \delta_{J,R} \, \delta_{R,W} \, \delta{m,t}
\end{split}
\end{equation}
we then have:
\begin{equation} \label{nrc_min_app}
\begin{split}
d\left( 1- \overline{G_\mathcal{A}(U_t)}^{NRC} \right)&=\sum_{(R,p,m),(W,s,t)} \delta_{R,W} \frac{\delta_{p,s}}{n_R} \, \frac{\delta_{p,s}}{d_R}+ \sum_{(R,p,m),(W,s,t)} \delta_{R,W} \frac{\delta_{m,t}}{n_R} \, \frac{\delta_{m,t}}{d_R} - \sum_{(R,p,m)} \frac{1}{n_R} \, \frac{1}{d_R} \\
&= \sum_R n_R\,d_R^2 \, \frac{1}{n_Rd_R} + \sum_R n_R^2 \, d_R \, \frac{1}{n_Rd_R} - \sum_R n_Rd_R \, \frac{1}{n_Rd_R}\\
&=\sum_J (d_J+n_J-1)=\sum_J d_J + \sum_J n_J - d_\mathcal{Z}
\end{split}
\end{equation}

\par Now, consider the family of Hamiltonians that satisfy NRC and are of the form $H=\oplus_J H_J$. This implies that each $J$-subspace is spanned by exactly $n_J\, d_J$ orthonormal energy eigenstates, i.e. $\ket{\phi_k} = \ket{\phi_{J,a}}$ with $J=1,\dots , \mathcal{Z}$, $a=1,\dots , n_J \, d_J$. \cref{alt_NRC} takes the form:
\begin{equation}
\begin{split}
d\left( 1- \overline{G_\mathcal{A}(U_t)}^{NRC} \right)=\sum_{J,a,b} \frac{\langle \rho_{J,a}^{n_J} , \rho_{J,b}^{n_J} \rangle^2}{n_J \, d_J}
+\sum_{J,a,b} \frac{\langle \rho_{J,a}^{d_J} , \rho_{J,b}^{d_J} \rangle^2}{n_J \, d_J}
- \frac{1}{2}\sum_{J,a} \frac{\langle \rho_{J,a}^{n_J} , \rho_{J,a}^{n_J} \rangle^2 + \, \langle \rho_{J,a}^{d_J} , \rho_{J,a}^{d_J} \rangle^2 }{n_J \, d_J} 
\end{split}
\end{equation}
Notice that $\rho_{J,a}^{n_J}$ are reduced density matrices, so $\lVert \rho_{J,a}^{n_J} \rVert_1 = 1$ and $\lVert \rho_{J,a}^{n_J} \rVert_2^2 = \lVert \rho_{J,a}^{d_J} \rVert_2^2 =: P_J^a \leq 1$. So:
\begin{equation} \label{ineqs}
\begin{split}
&\langle \rho_{J,a}^{n_J} , \rho_{J,b}^{n_J} \rangle \leq \lVert \rho_{J,a}^{n_J} \rVert_1 \, \lVert \rho_{J,b}^{n_J} \rVert_\infty \leq \lVert \rho_{J,b}^{n_J} \rVert_2 = \sqrt{P_J^b} \\
&\langle \rho_{J,a}^{d_J} , \rho_{J,b}^{d_J} \rangle \leq \sqrt{P_J^b}
\end{split}
\end{equation}
Also,
\begin{equation}
\begin{split}
&\sum_a \rho_{J,a}^{n_J} = \sum_a \Tr_{d_J} (\ket{\phi_{J,a}} \bra{\phi_{J,a}}) = \Tr_{d_J} ( \mathds{1}_{n_J \, d_J} ) = d_J \, \mathds{1}_{n_J} \\
&\sum_a \rho_{J,a}^{d_J} =  n_J \, \mathds{1}_{d_J}
\end{split}
\end{equation}
So:
\begin{equation}
\begin{split}
d\left( 1- \overline{G_\mathcal{A}(U_t)}^{NRC} \right) &\leq \sum_{J,a,b} \frac{\langle \rho_{J,a}^{n_J} , \rho_{J,b}^{n_J} \rangle \sqrt{P_J^b}}{n_J \, d_J} + \sum_{J,a,b} \frac{\langle \rho_{J,a}^{d_J} , \rho_{J,b}^{d_J} \rangle \sqrt{P_J^b}}{n_J \, d_J} - \sum_{J,a} \frac{{P_J^a}^2}{n_J \, d_J}\\
&= \sum_{J,b} \frac{\langle \mathds{1}_{n_J} , \rho_{J,b}^{n_J} \rangle \sqrt{P_J^b}}{n_J} +  \sum_{J,b} \frac{\langle \mathds{1}_{d_J} , \rho_{J,b}^{d_J} \rangle \sqrt{P_J^b}}{d_J} - \sum_{J,a} \frac{{P_J^a}^2}{n_J \, d_J} \\
&= \sum_{J,a} \left(\sqrt{P_J^a} \left(\frac{1}{n_J} + \frac{1}{d_J} \right) - \frac{{P_J^a}^2}{n_J \, d_J} \right) =: \sum_{J,a} f(P_J^a)
\end{split}
\end{equation}
Notice that
\begin{equation} \label{deriv}
\frac{\partial f}{\partial P_J^a} = \frac{n_J + d_J - 4 {P_J^a}^{3/2}}{2n_J d_J \sqrt{P_J^a}}
\end{equation}
Now, for all $J$'s for which $n_J,d_J \geq 2$, the partial derivative \cref{deriv} is non-negative $\forall$ $P_J^a \in (1/\min\{n_J,d_J\} , 1)$, so $f$ is maximized for $P_J^a =1$. Meanwhile, if either $n_J=1$ or $d_J=1$, the purity $P_J^a =1$. So:
\begin{equation}
d\left( 1- \overline{G_\mathcal{A}(U_t)}^{NRC} \right) \leq \sum_{J,a} f(\{P_J^a=1\})= \sum_{J,a} \left(\frac{1}{n_J}+\frac{1}{d_J}-\frac{1}{n_J d_J} \right) = \sum_J \left(n_J + d_J - 1 \right)
\end{equation}
Thus, the value \cref{nrc_min_app} is indeed the minimum. 

\subsection{Calculations for Example \ref{QRF_example}}
\begin{itemize}
    \item For the algebra $\mathcal{A}^1 = \langle \mathds{1}_{2S},\mathds{1}_2 \otimes \sigma_S^z , \mathds{1}_2 \otimes \sigma_S^x , \mathds{1}_2 \otimes \sigma_S^y \rangle$ we have $d_\mathcal{Z}=1, d_1=n_1=2$. It is convenient to work with the orthogonal basis $\{e_\alpha\}_{\alpha =1}^4 = \{\frac{\mathds{1}_{2S}}{2},\frac{\mathds{1}_2\otimes \sigma_S^z}{2},\frac{\mathds{1}_2 \otimes \sigma_S^x}{2}, \frac{\mathds{1}_2 \otimes \sigma_S^y}{2} \}$ of $\mathcal{A}^1$. Note that this is not the same with the basis in \cref{bases_formula}, but is unitarily related. We can, then, express the projectors $\mathbb{P}_{\mathcal{A}^1},\mathbb{P}_{{\mathcal{A}^1}^\prime}$ as
    \begin{equation} \label{projectors}
    \begin{split}
        &\mathbb{P}_{\mathcal{A}^1} (\cdot )=\sum_{\alpha =1}^4 \langle e_\alpha, (\cdot ) \rangle e_\alpha = \Tr( \cdot ) \frac{\mathds{1}_{2S}}{4} + \Tr( \Tr_2(  \cdot ) \vec{\sigma}_S ) \frac{\mathds{1}_2 \otimes \vec{\sigma}_S}{4} \\
        &\mathbb{P}_{{\mathcal{A}^1}^\prime} (\cdot ) = \sum_{\alpha=1}^4 e_\alpha (\cdot ) e_\alpha^\dagger = \frac{(\cdot )}{4} + \frac{\mathds{1}_2 \otimes \vec{\sigma}_S \, (\cdot ) \, \mathds{1}_2 \otimes \vec{\sigma}_S}{4}
    \end{split}        
    \end{equation}
    where we also used that the $e_\alpha$'s are already normalized, $\lVert e_\alpha \rVert_2=1$. In $R_1$ frame the dynamics $\mathcal{U}_t (\cdot )= \exp(it \, H) (\cdot ) \exp(-it \, H)$ are given by the Hamiltonian $H_{\overline{1}}=J_z \, \sigma_2^z \otimes \sigma_S^z + J_x \, \sigma_2^x \otimes \sigma_S^x + J_y \, \sigma_2^y \otimes \sigma_S^y$ with the eigenstates being the Bell states $\{\phi_k\}_{k=1}^4= \{\ket{\phi^{+}}, \ket{\phi^{-}}, \ket{\psi^{+}}, \ket{\psi^{-}} \}$. Writing out \cref{NRC} explicitly we have
    \begin{equation} \label{nrc_expression}
    \begin{split}
        \overline{G_{\mathcal{A}^1}(\mathcal{U}_t)}^{NRC}=1-\frac{1}{4} &\left(2\sum_{\substack{k,l=1\\k<l}}^4 \left\lVert \mathbb{P}_{\mathcal{A}^1} (\ketbra{\phi_k}{\phi_l}) \right\rVert_2^2 \, \left\langle \mathbb{P}_{{\mathcal{A}^1}^\prime} ( \ketbra{\phi_k}{\phi_k} ) , \mathbb{P}_{{\mathcal{A}^1}^\prime} ( \ketbra{\phi_l}{\phi_l} ) \right\rangle \right. +\\
        &\left. \hspace{20pt}+\, 2 \sum_{\substack{k,l=1\\k<l}}^4 \left\lVert \mathbb{P}_{{\mathcal{A}^1}^\prime} (\ketbra{\phi_k}{\phi_l}) \right\rVert_2^2 \, \left\langle \mathbb{P}_{\mathcal{A}^1} ( \ketbra{\phi_k}{\phi_k} ) , \mathbb{P}_{\mathcal{A}^1} ( \ketbra{\phi_l}{\phi_l} ) \right\rangle \, + \right.\\
        &\left.\hspace{20pt} +\sum_{k=1}^4 \left\lVert \mathbb{P}_{{\mathcal{A}^1}} (\ketbra{\phi_k}{\phi_k}) \right\rVert_2^2 \, \left\lVert \mathbb{P}_{{\mathcal{A}^1}^\prime} (\ketbra{\phi_k}{\phi_k}) \right\rVert_2^2 \right)  
    \end{split}
    \end{equation}
    where we used the fact that the terms with $k>l$ are equal to those with $k<l$. Substituting \cref{projectors} in \cref{nrc_expression} one obtains, in a rather tedious but straightforward manner, that $\overline{G_{\mathcal{A}^1}(\mathcal{U}_t)}^{NRC}=3/4$.
    \item The algebra $\mathcal{A}_i^{\vec{\eta}}=\langle \mathds{1}_{iS}, \mathds{1}_i\otimes \sigma_S^{\vec{\eta}}\rangle$ is Abelian and we have $d_\mathcal{Z}=2,d_1=d_2=1,n_1=n_2=2$. We work with the orthogonal basis $\{e_\alpha \}_{\alpha=1}^2=\{\frac{\mathds{1}_{iS}}{\sqrt{2}},\frac{\mathds{1}_{i} \otimes \sigma_S^{\vec{\eta}}}{\sqrt{2}}\}$. Then
    \begin{equation} \label{projectors2}
        \begin{split}
        &\mathbb{P}_{\mathcal{A}_i^{\vec{\eta}}} (\cdot )=\sum_{\alpha =1}^2 \left\langle \frac{e_\alpha}{\lVert e_\alpha \rVert_2} , (\cdot ) \right\rangle \frac{e_\alpha}{\lVert e_\alpha \rVert_2} = \Tr( \cdot ) \frac{\mathds{1}_{iS}}{4} + \Tr( \Tr_i(  \cdot ) \vec{\sigma}_S ) \frac{\mathds{1}_i \otimes \sigma_S^{\vec{\eta}}}{4} \\
        &\mathbb{P}_{{\mathcal{A}_i^{\vec{\eta}}}^\prime} (\cdot ) = \sum_{\alpha=1}^2 e_\alpha (\cdot ) e_\alpha^\dagger = \frac{(\cdot )}{2} + \frac{\mathds{1}_i \otimes \sigma_S^{\vec{\eta}} \, (\cdot ) \, \mathds{1}_i \otimes \sigma_S^{\vec{\eta}}}{2}
        \end{split}
    \end{equation}
    In $R_1$ frame the Hamiltonian is $H_{\overline{1}}=J_z \, \sigma_2^z \otimes \sigma_S^z + J_x \, \sigma_2^x \otimes \sigma_S^x + J_y \, \sigma_2^y \otimes \sigma_S^y$ with the eigenstates being the Bell states $\{\phi_k\}_{k=1}^4= \{\ket{\phi^{+}}, \ket{\phi^{-}}, \ket{\psi^{+}}, \ket{\psi^{-}} \}$, while in $R_2$ frame the Hamiltonian is $  H_{\overline{2}} =J_z \, \mathds{1}_1 \otimes \sigma_S^z + J_x \, \sigma_1^x \otimes \mathds{1}_S - J_y \, \sigma_1^x \otimes \sigma_S^z$ with the eigenstates being $\{ \ket{+0}, \ket{+1},\ket{-0},\ket{-1}\}$, where $\ket{\pm}\equiv \frac{1}{\sqrt{2}} (\ket{0} \pm \ket{1} )$ are the eigenstates of $\sigma_x$. Using the analogous expression of \cref{nrc_expression} and substituting \cref{projectors2} for the two frames one obtains \cref{lta_R1,lta_R2}.
\end{itemize}

\subsection{Gradient Descent for Test of Conjecture}\label{grad_appendix}

The distinguished basis corresponding to $\mathcal{A}$ consists of vectors $\mathbb{B}=\left\{\ket{p_J}\otimes\ket{a_J}\right\}_{p_J=1, \ a_J =1}^{n_J,\  d_J}$ and is unique up to local unitaries on the virtual bipartitions in each $J$-block of \cref{structure_H}. In this basis, the eigenstates of the conjectured Hamiltonian (see \hyperlink{conj}{Conjecture}) are the "canonical" basis vectors of the full Hilbert space. For an arbitrary NRC Hamiltonian $H$, the eigenbasis can be expressed as a rotation of $\mathbb{B}$ by a unitary $U$, so that the one-dimensional eigenprojectors are given as $U\Pi_kU^\dagger$, where $\Pi_k\equiv \ketbra{p_J,a_J}{p_J,a_J}$ are the projectors corresponding to $\mathbb{B}$.

\par In \cref{num_evidence} we utilize a gradient descent algorithm by Abrudan et al. \cite{abrudan_steepest_2008} in order to numerically minimize \cref{NRC} over all possible eigenbases for a given algebra $\mathcal{A}$. For a given $\mathcal{A}$, the $\mathcal{A}$-OTOC NRC LTA is a function $\overline{G_{\mathcal{A}}(\mathcal{U}_t )}^{NRC}(U): U(d) \rightarrow \mathbb{R}$ over the manifold $U(d)$ of $d\times d$ unitaries, which we can embed in a Euclidean space $\mathbb{C}^{d\times d}$ with inner product $\langle A,B \rangle_{\mathbb{C}^{d\times d}} \coloneqq \mathscr{R}(\Tr(AB^\dagger))$, where $\mathscr{R}$ denotes the real part. Note that $\mathcal{U}_t (\cdot ) = \exp(it \, H) (\cdot ) \exp(-it \, H)$ is the unitary Hamiltonian evolution in the Heisenberg picture, while $U$ is the unitary that represents the transformation of the eigenbasis of $H$ to $\mathbb{B}$. 
\par Starting from a random unitary $U_0$ we move on $U(d)$ along the geodesic with the steepest descent for $\overline{G_{\mathcal{A}}(\mathcal{U}_t )}^{NRC}(U)$ via $U_1 \equiv \exp (-\mu \, G_{U_0} U_{0})$. Here, $\mu \in \mathbb{R}^+$ is a dynamic step size adjusted to improve the rate of convergence \cite{abrudan_steepest_2008} and $G_{U_0}\in T_{\mathds{1}}U(d)$ is the Riemannian gradient of $\overline{G_{\mathcal{A}}(\mathcal{U}_t )}^{NRC}(U)$ at $U_{0}$ translated to the tangent space at the identity. Choosing the inner product $\langle X,Y \rangle_U \coloneqq \frac{1}{2} \mathscr{R}(\Tr(XY^\dagger))$ for the tangent space of $U(d)$ at $U$, we have $G_U\equiv\Gamma_U U^\dagger- U \Gamma_U^\dagger$ \cite{abrudan_steepest_2008}, where $\Gamma_U \coloneqq \nabla_{U^*}\overline{G_{\mathcal{A}}(\mathcal{U}_t)}^{NRC}$ is the standard Euclidean gradient on $\mathbb{C}^{n\times n}$. Using the first expression in \cref{nrcp} for the one-dimensional eigenprojectors $U\Pi_k U^\dagger$, the function $\overline{G_{\mathcal{A}}(\mathcal{U} )}^{NRC}(U)$ is written explicitly as
\begin{equation} \label{nrc_U}
\begin{split}
    \overline{G_\mathcal{A} (\mathcal{U}_t)}^{NRC}(U)&=1-\frac{1}{d} \sum_{\alpha=1}^{d(\mathcal{A})} \sum_{\gamma=1}^{d(\mathcal{A}^\prime)} \left(\sum_{k,l=1}^d \Tr(e_\alpha U\Pi_k U^\dagger f_\gamma U\Pi_k U^\dagger e_\alpha^\dagger U\Pi_l U^\dagger f_\gamma^\dagger U\Pi_l U^\dagger)+\right.\\
    &\hspace{100pt}+\sum_{k,l=1}^d \Tr(e_\alpha U\Pi_k U^\dagger f_\gamma U\Pi_l U^\dagger e_\alpha^\dagger U\Pi_l U^\dagger f_\gamma^\dagger U\Pi_k U^\dagger)-\\
    &\hspace{100pt}\left.-\sum_{k=1}^d\Tr(e_\alpha U\Pi_k U^\dagger f_\gamma U\Pi_k U^\dagger e_\alpha^\dagger U\Pi_k U^\dagger f_\gamma^\dagger U\Pi_k U^\dagger)\right)
\end{split}
\end{equation}
For the Euclidean gradient of a function $f(U):\mathbb{C}^{d\times d} \rightarrow \mathbb{R}$, we have $\delta f = \langle \nabla_U f , \delta U \rangle_{\mathbb{C}^{d\times d}}+\langle \nabla_{U^*} f , \delta U^* \rangle_{\mathbb{C}^{d\times d}}=2 \mathscr{R}\left(\Tr(\left(\nabla_{U^*}f\right)^T \delta U)\right)$, where $A^T$ denotes the matrix transpose and we used that $(\nabla_U f)^*=\nabla_{U^*} f$. Performing the variation in \cref{nrc_U} and comparing with $\delta \overline{G_{\mathcal{A}}(\mathcal{U}_t )}^{NRC} = 2 \mathscr{R}\left(\Tr(\Gamma_U^T \delta U)\right)$ we find, after some algebraic manipulation, that
\begin{equation}
\begin{split}
\Gamma_U &= 2\sum_{kl} \bigg((1-\delta_{kl}/2) \\
    &\hspace{45pt}\left(\left[ \sum_\gamma \left(\mathcal{P}_{\mathcal{A}^\prime}\left(U\Pi_lU^\dagger f_\gamma U\Pi_l U^\dagger \right)U\Pi_kU^\dagger f_\gamma^\dagger \right)+
    \sum_\alpha \mathcal{P}_\mathcal{A}\left(U\Pi_lU^\dagger e_\alpha U\Pi_l U^\dagger \right)U\Pi_kU^\dagger e_\alpha^\dagger \right] + \text{h.c.} \right) U\Pi_k\bigg) .
\end{split}
\end{equation}
where h.c. denotes the hermitian conjugate of the expression inside the brackets.
\par We iteratively update the unitary $U_{k+1}=\exp(-\mu_k G_{U_k} U_k )$ until the convergence condition $\abs{\overline{G_\mathcal{A} (\mathcal{U}_t)}^{NRC}(U_{k+1})-\overline{G_\mathcal{A} (\mathcal{U}_t)}^{NRC}(U_k)}<\epsilon$ is met, where $\epsilon$ is a tolerance set to $\epsilon=10^{-8}$.



\end{document}